\documentclass[
  aps,
  prl,
  reprint,
  superscriptaddress,
  nofootinbib,
  longbibliography,
  floatfix
]{revtex4-2}

\makeatletter
\newcommand*{\rom}[1]{\expandafter\@slowromancap\romannumeral #1@}
\makeatother
\usepackage{graphicx}
\graphicspath{{img/}{./}} 
\usepackage{amsmath, amssymb, amsfonts, mathtools, bm}
\usepackage[colorlinks=true,linkcolor=blue,citecolor=blue,urlcolor=blue]{hyperref}
\usepackage{natbib}
\usepackage{float}
\usepackage[T1]{fontenc}
\usepackage{listings}
\usepackage{xcolor}
\usepackage{enumitem}

\usepackage{ragged2e}

\begin{document}

\title{General-Purpose Inverse Design of Heterogeneous Finite-Sized Assemblies}

\author{Livia A. J. Guttieres}
\affiliation{School of Engineering and Applied Sciences, Harvard University, Cambridge, Massachusetts 02138, USA}
\author{Ryan K. Krueger}
\affiliation{School of Engineering and Applied Sciences, Harvard University, Cambridge, Massachusetts 02138, USA}
\author{Remi Drolet}
\affiliation{School of Engineering and Applied Sciences, Harvard University, Cambridge, Massachusetts 02138, USA}
\author{Michael P. Brenner}
\affiliation{School of Engineering and Applied Sciences, Harvard University, Cambridge, Massachusetts 02138, USA}
\affiliation{Department of Physics, Harvard University, Cambridge, Massachusetts 02138, USA}

\date{\today}

\begin{abstract}
Designing heterogeneous, self-assembling systems is a central challenge in soft matter and biology.
We present a framework that uses gradient-based optimization to invert an analytical yield calculation, tuning systems toward target equilibrium yields.
We design systems ranging from simple dimers to temperature-controlled shells to polymerizing systems, achieving precise control of self- and non-self-limiting assemblies.
By operating directly on closed-form calculations, our framework bypasses trajectory-based instabilities and enables efficient optimization in  challenging regimes.
\end{abstract}

\maketitle

The self-assembly of target structures from heterogeneous, interacting building blocks underlies a broad range of biological and synthetic systems~\cite{hagan2008review, mcmanus2016physics, grzybowski2009self, lutz2019sequence}.
Designing interactions that produce target structures with high yield is therefore a central problem across soft matter physics, materials science, and biophysics~\cite{jain2013inverse, lai2012principles, sanchez2012hydrophobic, law2013self, sacanna2011shape, jacobs2015rational}.
Simulations of patchy particles show how anisotropic binding patches can drive robust formation of monodisperse clusters under reversible dynamics~\cite{wilber2007patchy}.
The standard inverse design approach is to invert a forward model, ranging from molecular dynamics simulations to analytical calculations of the assembly yield or free energy~\cite{lindquist2016communication, jadrich2017probabilistic}.

However, forward models vary in accuracy and cost; detailed simulations are often prohibitively expensive for design while analytical calculations typically assume simplified systems such as isotropically interacting spheres~\cite{klein2018physical, geng2019engineering, van2015digital, hormoz2011design}.
While recent work~\cite{king2024programming, krueger2024tuning} advanced inverse design by directly optimizing through molecular dynamics simulations, target behaviors remain limited by (i) computational complexity, (ii) challenges in computing rare-event statistics, and (iii) discontinuities in computing discrete variables (e.g., counting candidate structures).
Inverse design methods therefore face a tradeoff: simulations capture entropy and anisotropy but are hampered by instability and sampling demands, while analytical approaches are efficient yet neglect key assembly effects~\cite{klein2018physical, meng2010free, zeravcic2017colloquium}.

Here, we adapt a recent analytical framework for computing the grand-canonical assembly yield of heterogeneous building blocks~\cite{curatolo2023toolbox} to enable the design of complex self-assembling structures.
The framework explicitly models translational, rotational, and vibrational entropic contributions and concentration dependence of arbitrarily shaped, anisotropically interacting building blocks.
It first computes partition functions for a set of candidate assemblies, and then  numerically solves for their concentrations via a self-consistent system of equations.
We introduce an end-to-end framework to differentiate this calculation, enabling flexible optimization of arbitrary control parameters (e.g. input monomer concentrations, interaction parameters).
This enables (i) the design of anisotropic systems that capture entropic contributions, validated against but independent of canonical ensemble simulation, and (ii) the tuning of concentration dependence, which is inaccessible in differentiable MD.

\begin{figure*}
    \centering
    \includegraphics[width=0.86\textwidth]{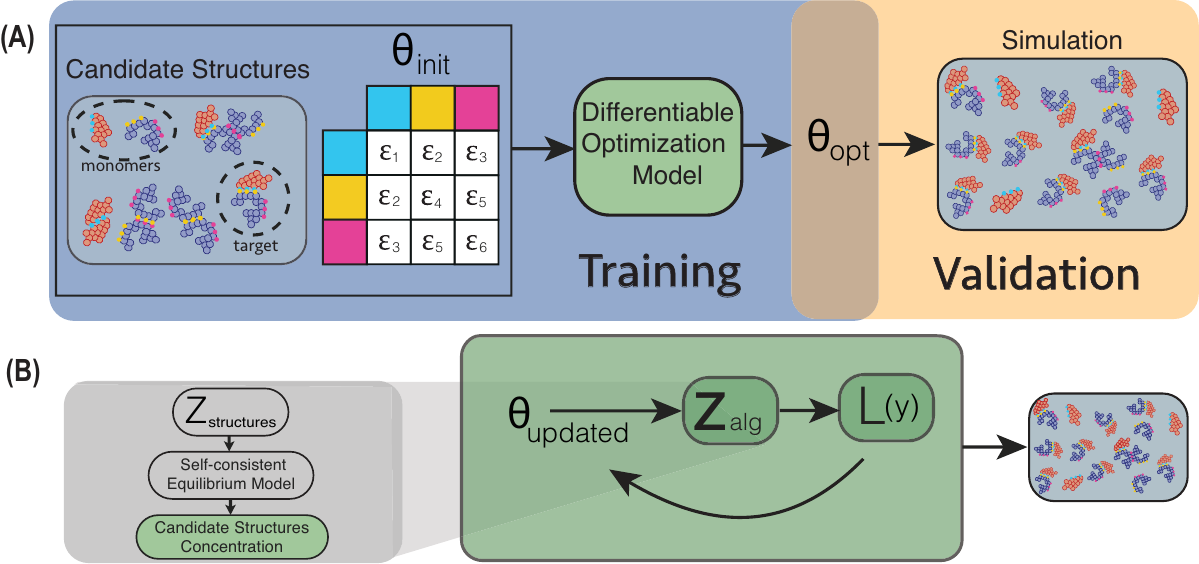}
    \caption{
    \textbf{Framework for designing heterogeneous, finite-sized assemblies.}
    \textbf{(A)}
    Optimization and validation pipeline.
    Candidate structures (e.g. stoichiometry, ground states) and initial parameters $\theta_{\mathrm{init}}$ inform a closed-form calculation of equilibrium yields.
    We differentiate these yields to update parameters via gradient descent, optimizing a user-defined objective.
    Optimized parameters $\theta_{\mathrm{opt}}$ are validated via MD simulations.
    \textbf{(B)} Differentiable optimization model.
    Equilibrium yields are computed analytically by (i) calculating structural partition functions, and (ii) mapping these to concentrations by numerically solving a self-consistent system of equations.
    The resulting yields are used to evaluate the objective function, $\mathcal{L}$.
    This entire procedure is implemented in a differentiable form, enabling automatic differentiation and gradient-based updates of $\theta$.
    }
    \label{fig:optimization-schematic}
\end{figure*}

We first illustrate our approach using a minimal dimer system at finite concentration.
We then examine two broader classes of assemblies: (i) closed-shell structures that assemble and disassemble under controlled conditions, and (ii) polymerizing systems that are non-self-limiting in principle but exhibit a target size distribution.
For the shells, we use multi-ensemble optimization to achieve controlled disassembly within a target temperature range, inspired by cellular cargo delivery.
For the polymers, the challenge is more severe: the dominant off-targets cannot be enumerated or approximated \emph{a priori}.
We address this with an auxiliary objective based on a generalized mass action constraint~\cite{hagan2021selflimiting} that imposes concentration-dependent penalties on overgrowth.
We extend the known single-species form to multi-species systems through a novel approximation, enabling selective yield targeting in unbounded growth regimes.
This objective substantially improves agreement with simulation, underscoring the our framework's flexibility to incorporate richer theories of self-assembly.
We optimize temperature and interaction parameters for dimers and shells.
For polymers, we also tune monomer concentrations, as highly non-stoichiometric conditions can mitigate yield catastrophes~\cite{murugan2015undesired}.

Together, these contributions advance heterogeneous self-assembly from a descriptive theory to an actionable design tool, extending beyond differentiable MD through efficient, stable equilibrium design that incorporates entropy, anisotropy, and concentration dependence.

\textit{Optimization Framework.---}
Following Curatolo et al.~\cite{curatolo2023toolbox}, we consider a system composed of $N$ rigid building blocks with short-range interactions.
Each target cluster $s$ is characterized by a potential energy $E_s(\mathbf{q}, \bm{\phi})$ that depends on the translational and rotational degrees of freedom of the constituent monomers ($\mathbf{q}$ and $\bm{\phi}$, respectively).
For rigid clusters, Curatolo et al. introduce a tractable approximation to the configurational partition function $Z_s$ for each cluster $s$ via (i) a change of variables to global cluster translations, rotations, and internal vibrational modes, and (ii) the assumption that the thermal energy is small relative to the potential energy (see Supplemental Material~\cite{supplement}, Sec.A.2).
The resulting approximation for $Z_s$ is as follows:
\begin{align}
Z_s &\approx e^{-\beta E_0} \times V \times \frac{\widetilde{J}}{\sigma_s} \times
\prod_{i=1}^{6N_s - 6} \sqrt{\frac{2\pi}{\beta \omega_i^2}} \nonumber\\
&\equiv e^{-\beta E_0} \times Z_s^{\text{trans}} \times Z_s^{\text{rot}} \times Z_s^{\text{vib}} .
\label{eq:part_fn}
\end{align}
where Equation \ref{eq:part_fn} is the approximation introduced in Ref.~\cite{curatolo2023toolbox}, with $\Omega_s$ denoting the region of configuration space where $s$ is defined, $\sigma_s$ is a symmetry number accounting for indistinguishable configurations, and $\beta = 1 / k_B T$ is the inverse thermal energy where $k_B$ is the Boltzmann constant and $T$ is the temperature (see SM, Sec.A for details).
For the approximation described by Equation \ref{eq:part_fn}, $E_0$ is the ground state energy of $s$, $\omega_i^2$ are the nonzero eigenvalues of the Hessian of the ground state energy with respect to the vibrational modes, and $\widetilde{J}$ is the integral of the Jacobian over global rotations.
$Z_{s}^{\text{trans}}$, $Z_{s}^{\text{rot}}$, and $Z_{s}^{\text{vib}}$ denote the translational, rotational, and vibrational partition functions, respectively.

Given the partition functions $\{Z_s\}$, Curatolo et al. introduce a numerical scheme for computing the equilibrium concentrations of each cluster, $\{c_s\}$.
Specifically, $\{c_s\}$ are the solution to a coupled nonlinear system of equations, consisting of:
\begin{itemize}[noitemsep,topsep=2pt,parsep=0pt,partopsep=0pt,leftmargin=*]
  \item A conservation law for each monomer species $\alpha$:
  \begin{align}
  \sum_s N_{s,\alpha} c_s &= c_\alpha^{\text{tot}},
  \label{eq:selfconsistent2}
  \end{align}
  where $c_\alpha^{\text{tot}}$ is the total concentration of monomer $\alpha$ and
  $N_{s,\alpha}$ is the number of copies of monomer $\alpha$ in structure $s$.

  \item A mass-action constraint for each non-monomeric cluster $s$:
  \begin{align}
  V c_s \prod_\alpha c_\alpha^{N_{s,\alpha}} &= Z_s \prod_\alpha Z_\alpha^{-N_{s,\alpha}},
  \label{eq:selfconsistent1}
  \end{align}
  where $c_\alpha$ and $Z_\alpha$ denote the concentration and partition function of monomer $\alpha$, respectively.
\end{itemize}
Note that monomers are valid equilibrium assemblies, with $c_{\alpha}^{\text{tot}}$ denoting the input concentration of monomer $\alpha$ while $c_{\alpha}$ denotes the equilibrium concentration of the monomeric structure $s_{\alpha}$.
The final assembly yields ${Y_s}$ are obtained by normalizing the equilibrium concentrations.
Taken together, Equation \ref{eq:part_fn}, Equation \ref{eq:selfconsistent2}, and Equation \ref{eq:selfconsistent1} define a complete calculation for computing the assembly yield of a set of rigid candidate assemblies ${s}$ given (i) their ground state configurations, (ii) an energy function, (iii) the input concentrations of monomeric species, and (iv) temperature.
\begin{figure}
    \centering
    \includegraphics[width=0.47\textwidth]{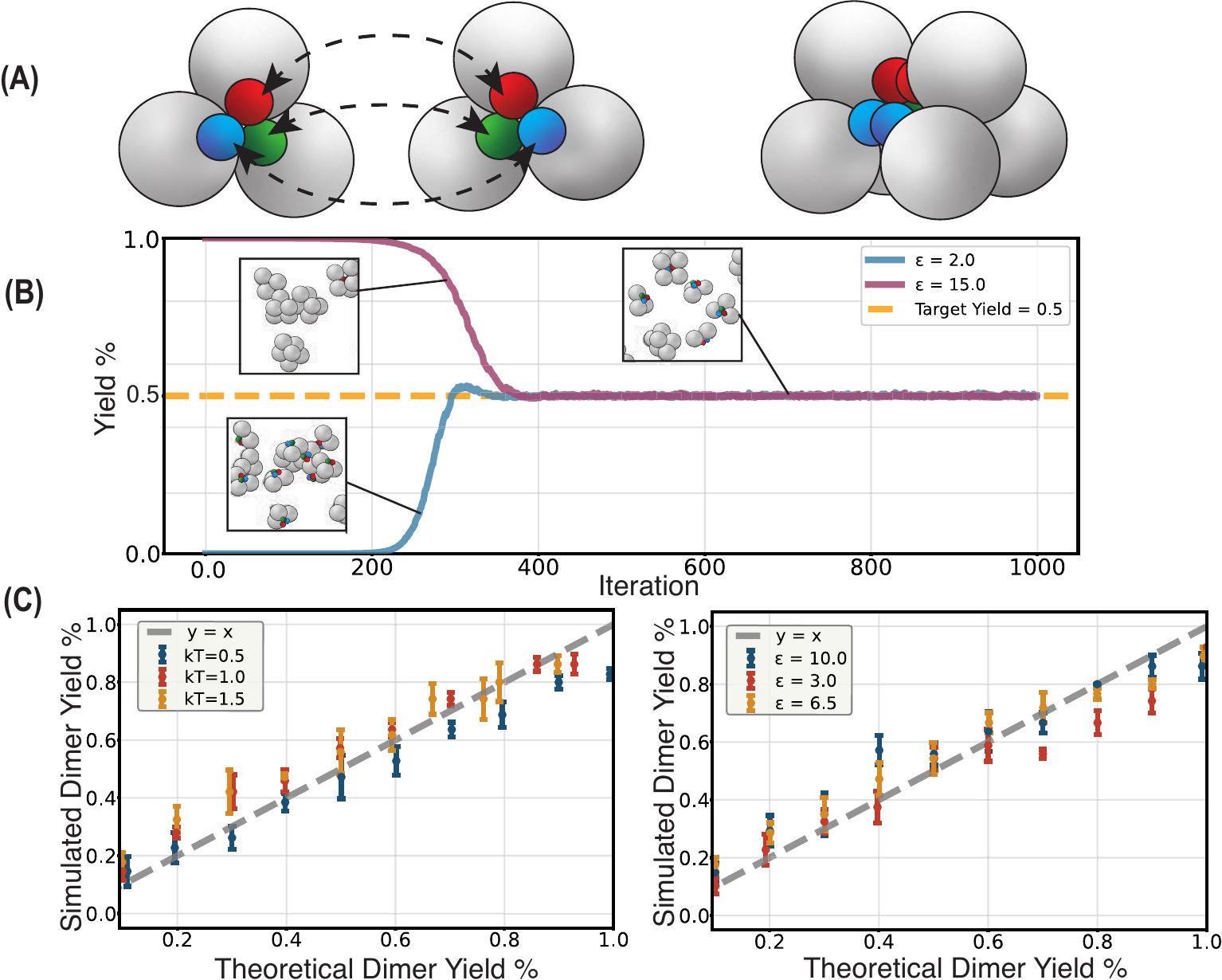}
    \caption{
    \textbf{Dimer optimization and validation.}
    \textbf{(A)} Schematic of two enantiomeric monomers, each with three distinct patches.
    Identical patch types interact, yielding a single target dimer.
    \textbf{(B)} Optimization convergence toward target equilibrium yield $0.5$.
    Purple and blue curves denote initialization with with strong (large $\epsilon$) and weak (small $\epsilon$) attraction, respectively.
    Insets show representative MD snapshots.
    \textbf{(C)} Theoretical vs. MD yields across independent optimizations.
    Left: $\epsilon$ fixed, temperature optimized.
    Right: temperature fixed, $\epsilon$ optimized.
    Most final theoretical yields are within $1\%$ of the target.
    }
\label{fig:dimerfigure}
\end{figure}
To transform yield prediction into a design framework, we introduce a procedure for directly differentiating the assembly yield calculation described above.
For arbitrary continuous control parameters $\bm{\theta}$ (e.g. temperature, energy function parameters, monomer concentrations), the goal is to efficiently and precisely compute $\frac{dY}{d\bm{\theta}}$ where $Y = \{Y_s\}$ denotes the assembly yields given $\bm{\theta}$.
Given such a scheme, one could flexibly optimize $\bm{\theta}$ to minimize an arbitrary loss function defined over these assembly yields using gradient-based optimization.

We compute the total derivative $\frac{dY}{d\bm{\theta}} = \frac{\partial Y}{\partial Z}\cdot\frac{dZ}{d\bm{\theta}}$ by combining automatic differentiation of the partition function calculation with implicit differentiation of the fixed-point system defined by Eqs.~\ref{eq:selfconsistent2}--\ref{eq:selfconsistent1}~\cite{blondel2021implicit}. This avoids backpropagating through the iterative solver and bypasses the expensive sampling required for $\widetilde{J}$ (see End Matter). We implement the pipeline in JAX~\cite{jax2018github}, enabling optimization of $\bm{\theta}$ via gradient descent on a differentiable objective $\mathcal{L}(Y_{\bm{\theta}})$, where $Y_{\bm{\theta}}$ denotes the yields for $\bm{\theta}$.

To validate optimized parameters $\bm{\theta}_{\text{opt}}$, we performed molecular dynamics simulations.
Importantly, these MD simulations use the canonical ensemble with fixed particle number, whereas our optimization is formulated in a grand-canonical setting; their close quantitative agreement establishes that our framework transfers robustly to finite-sized canonical systems akin to experiments.

We evaluate our optimization framework across three representative test cases that span self-limiting and non-self-limiting behavior, as well as varying degrees of structural competition.
In each case, we compare predicted equilibrium yields against molecular dynamics simulations using the optimized parameters.

\textit{Case 1: Dimer.---}
First, we consider the toy dimer system introduced in Ref. \cite{curatolo2023toolbox}.
This system is composed of two enantiomeric rigid bodies, each composed of three spheres, and every sphere having a small colored patch that binds to like-colored patches via a Lennard-Jones potential (Figure \ref{fig:dimerfigure}A).
This minimal dimer system captures the essential physics of pairwise binding motifs in magnetic handshake materials and programmable colloids, where design tunes reversible, selective binding.
In magnetic handshake materials, interaction strength and selectivity are tuned via dipole spacing, strength, and geometry~\cite{niu2019magnetic, fenley2025hierarchical, du2022programming}, while in colloidal systems similar control is achieved through temperature, pH, depletant concentration, and ligand density~\cite{torquato2009inverse, jain2014perspective, dijkstra2021predictive}.

We first performed optimizations of the well depth, $\epsilon$, of each Lennard-Jones potential.
We used a target yield of $0.5$ which provides a stringent benchmark, as it requires the algorithm to tune the system to an intermediate state rather than trivially favoring either complete binding or complete dissociation.
Starting from two distinct initial conditions with either weak ($\epsilon = 2$) or strong ($\epsilon = 15$) interactions, the yield converged smoothly toward the target value in both cases (Figure \ref{fig:dimerfigure}B).

We further optimized either temperature (fixed $\epsilon$) or interaction strength (fixed $T$) across a range of target yields, with simulated yields closely matching targets in both cases (Figure \ref{fig:dimerfigure}C).

\textit{Case 2: Octahedral Shell.---}
We consider six monomers with two selectively-binding patch types~\cite{krueger2024tuning} that assemble into an octahedral shell. This is motivated by protein~\cite{bhaskar2017engineering, dowling2025hierarchical} and nucleic-acid~\cite{chandrasekaran2016dna, zhang2017placing, hao2014construction} cages, where switch-like assembly is tuned via solution conditions (temperature, pH, ionic strength) without altering the shape.

Beyond yield control, switch-like behavior requires a single parameter set to maximize yield at one condition and minimize it at another. Unlike Ref.~\cite{krueger2024tuning}, which disrupted pre-assembled shells, we target full assembly--disassembly, beyond the practical reach of simulation-based design given the long timescales of both processes.

\begin{figure}
    \centering
    \includegraphics[width=0.43\textwidth]{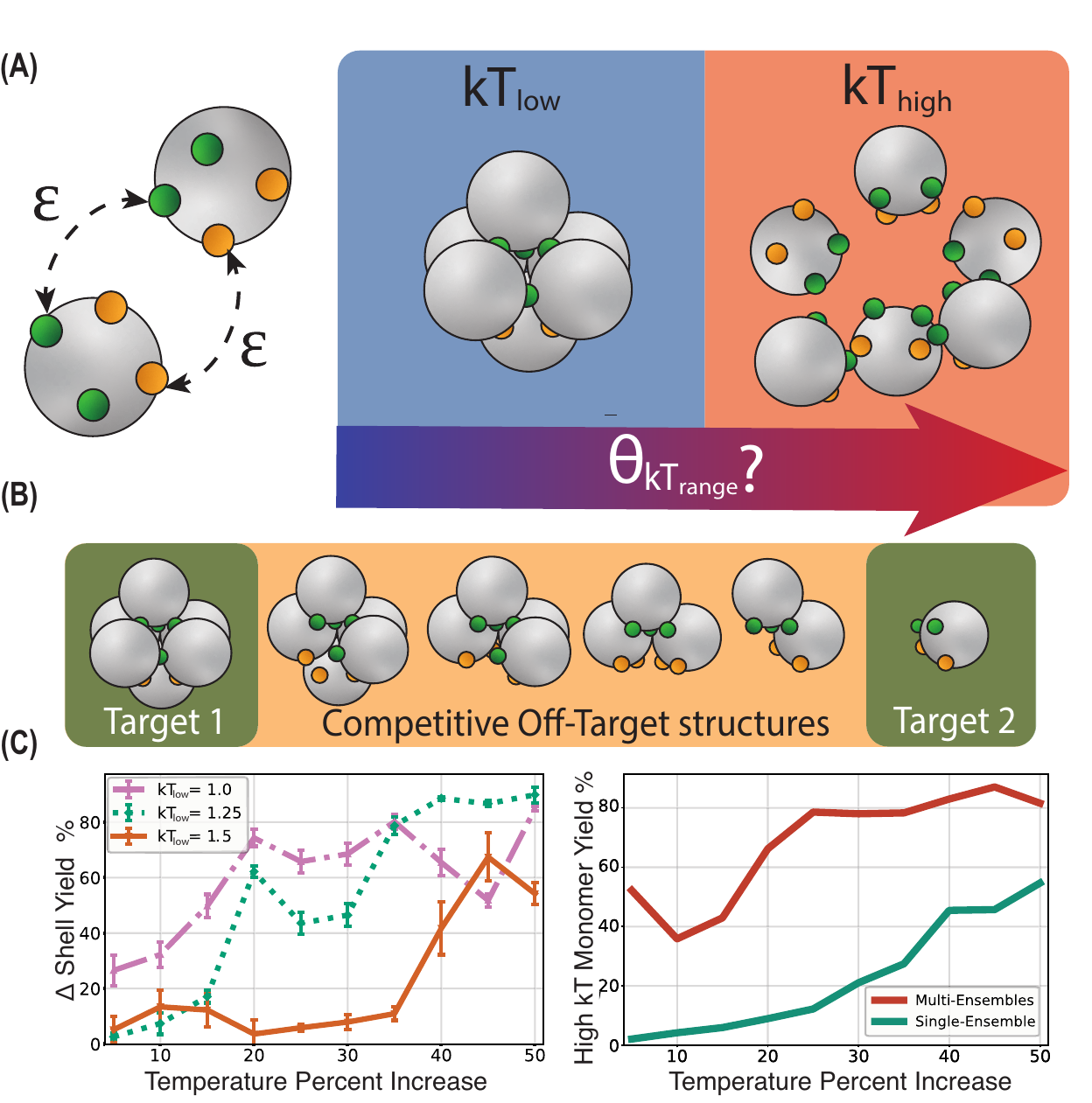}
    \caption{
    \textbf{Temperature-dependent control of shell assembly.}
    \textbf{(A)} System of patchy monomers with two patch types (green and yellow) that self-assemble into a closed shell.
    The optimization goal is finding parameters that produce switch-like behavior, favoring assembly at $kT_{\text{low}}$ and disassembly at $kT_{\text{high}}$.
    The red arrow highlights the challenge of maximizing yield contrast across this temperature range.
    \textbf{(B)} Modeled assembly outcomes: the fully assembled shell, a variety of incomplete off-target structures, and free monomers.
    \textbf{(C)} Optimization results.
    \emph{Left:} Absolute difference in simulated shell yields between $kT_{\text{low}}$ and $kT_{\text{high}}$ as a function of the percentage increase of $kT_{\text{high}}$ relative to $kT_{\text{low}}$, computed using the optimized parameters.
    The legend indicates the fixed baseline  ($kT_{\text{low}}$).
    \emph{Right:} Simulated monomer yields at $kT_{\text{high}}$ for two strategies: optimizing only for assembly at $kT_{\text{low}}$ (green) versus simultaneous multi-ensemble optimization (red).
    This demonstrates the necessity of multi-ensemble optimization for temperature-controlled assembly.
    }
    \label{fig:shell_optimization}
\end{figure}

This problem is more challenging than the dimer case in two ways. First, the space of off-target intermediates is intractable to enumerate, so we approximate it by a representative structure at each cluster size spanning fully assembled shells to free monomers (Figure \ref{fig:shell_optimization}B). Second, the loss must evaluate two ensembles simultaneously, balancing stability and disassembly (see SM).

Here, we focus on temperature as the environmental control variable, and the strengths of the Lennard-Jones interactions as the free parameters.
At low temperature ($kT_{low}$), the objective is to favor complete shell assembly, whereas at high temperature ($kT_{high}$), the objective is to favor shell disassembly (Figure \ref{fig:shell_optimization}A).
For a given value of $kT_{low}$, we perform optimizations across a range of higher temperatures.
Figure \ref{fig:shell_optimization}C (left) shows the resulting absolute difference in yield between the low and high temperatures using the optimized interaction parameters.
As the temperature gap increases, the optimizer is able to achieve a larger difference in yield, demonstrating increasingly precise control over the assembly process.
Importantly, even for relatively small temperature differences, the framework achieves appreciable changes in yield, highlighting both the sensitivity of the system and the accuracy of the optimization.

\begin{figure}
\centering
  \includegraphics[width=0.46\textwidth]{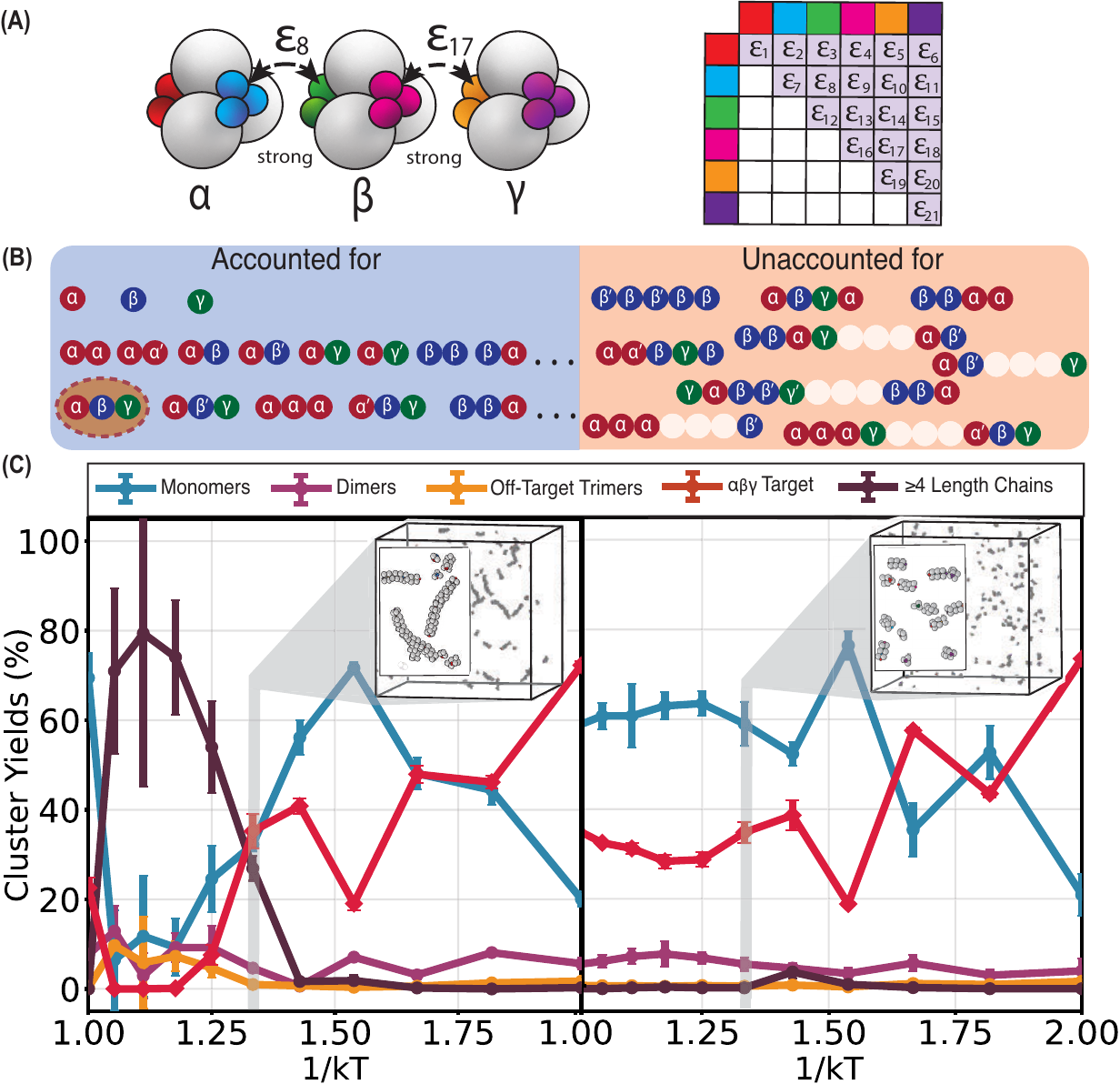}
\caption{
\textbf{Controlling polymer growth through mass action regularization.}
\textbf{(A)} Schematic of the polymerizing system.
We depict the target $\alpha\beta\gamma$ trimer with the two strong interactions that define the desired assembly highlighted.
The $6 \times 6$ interaction matrix $\varepsilon_{ij}$ describes all pairwise interactions between the six patch types across the three species ($\alpha$, $\beta$, and $\gamma$).
\textbf{(B)} Non-self-limiting polymerization challenge.
All monomers, dimers, and trimers can be explicitly included in the optimization (left).
However, longer chains ($n \geq 4$) grows combinatorially and cannot be exhaustively enumerated (right), making it intractable to directly account for every off-target structure in the analytical calculation.
\textbf{(C)} Simulated yield distributions by cluster size as a function of inverse temperature ($1 / kT$).
Parameters were optimized without (left) and with (right) mass action regularization.
Callouts depict representative snapshots at $kT = 0.75$.
}
\label{fig:polymer_comparison}
\end{figure}

Finally, we assess the importance of explicitly considering both ensembles in the optimization. Figure \ref{fig:shell_optimization}C (right) compares optimizations where the objective includes only the low-temperature ensemble versus both low- and high-temperature ensembles.
When only the goal of maximizing yield at low-temperature is considered in the optimization, the resulting parameters fail to reduce high-temperature assembly, leaving the system substantially resistant to disassembly.
In contrast, incorporating both ensembles into the loss function enables the optimizer to find parameters that stabilize the shell at $kT_{low}$ while driving near-complete disassembly at $kT_{high}$.
This result underscores the necessity of multi-ensemble optimization for achieving condition-dependent assembly and disassembly.

\textit{Case 3: Polymerization.---}
As a final example, we consider a polymerizing system in which chains can grow to arbitrary length.

This case is motivated by filament-forming biological and synthetic systems, such as microtubules~\cite{ndlec1997self, aranson2006theory}, actin filaments~\cite{alberts2002self, guo2009self}, and DNA-tile nanotubes~\cite{rothemund2004design, liu2004dna}, where unbounded growth competes with the formation of finite functional assemblies.
In these systems, filament length distributions are strongly regulated by monomer concentration and interaction energetics, making concentration an essential design variable rather than a fixed experimental constraint~\cite{hagan2021selflimiting}.

We consider three monomer types ($\alpha$, $\beta$, $\gamma$) with bidirectional attractive sites that link sequentially into chains. The design goal is to favor a specific finite trimer ($\alpha\beta\gamma$, Figure \ref{fig:polymer_comparison}A); following Ref.~\cite{murugan2015undesired}, we optimize both interaction strengths and input monomer concentrations.

This problem is significantly more challenging because growth is non-self-limiting: it is intractable to enumerate all off-target chains of increasing length (Figure \ref{fig:polymer_comparison}B). Approximating the space by a representative set of smaller off-targets, as in the shell and dimer cases, fails to capture thermodynamic competition from unbounded growth---simulations show longer chains emerging with lengths reaching up to 147, never explicitly prohibited by the optimization. Note that because polymer assembly is diffusion-limited at our dilute concentrations, finite-time MD yields should be interpreted as conservative lower bounds on the equilibrium values (see SM, Sec.F).

We instead add an auxiliary loss term, derived from a multi-species generalization of equilibrium mass action theory~\cite{hagan2021selflimiting}, that penalizes growth at the immediate overgrowth step ($n=4$) without requiring enumeration of longer chains (see End Matter). With this penalty, the optimizer favors the target trimer and suppresses longer polymers; without it, uncontrolled growth dominates at intermediate temperatures (Fig.~\ref{fig:polymer_comparison}C).

\textit{Discussion.---}
In this work we introduced a differentiable framework for optimizing self-assembling systems of heterogeneous building blocks.
By inverting an analytical yield calculation, our approach enables gradient-based tuning of physical parameters for target equilibrium behaviors.
Across case studies, we demonstrated flexibility: from simple two-component systems, to multi-ensemble optimization for temperature-dependent control, to non-self-limiting polymerizing systems, where growth must be managed by incorporating physical priors.
Simulations confirmed that our optimized parameters reliably yield the desired assembly outcomes.

The control parameters optimized here correspond to experimentally accessible variables in many self-assembling platforms: effective interaction strengths in magnetic handshake materials~\cite{niu2019magnetic, fenley2025hierarchical, du2022programming} and programmable colloids~\cite{torquato2009inverse, jain2014perspective, dijkstra2021predictive}, subunit concentrations in polymerizing and shell assemblies~\cite{hagan2021selflimiting, kushner1969self}, and environmental variables like temperature, pH, and ionic strength.
These parameters are routinely tuned experimentally and can modulate assembly behavior without altering the target geometry.

Several challenges remain. Our framework accounts for unenumerated states using representative off-targets or physical priors such as mass-action constraints; where neither is feasible, one could iteratively sample off-targets by simulation. Extending the framework to parameters that modify particle shape, bonding topology, or parameter-dependent ground states is also promising, achievable via a similar simulation-in-the-loop strategy~\cite{schoenholz2020jax, blondel2021implicit}. Where physical models or energy functions are uncertain, the same framework could fit them directly to experimental data, e.g., the Protein Data Bank~\cite{berman2000protein}.

We note that this equilibrium-based approach can help mitigate kinetic traps. By maximizing the target's equilibrium probability relative to off-targets, we deepen its energy well, which can disfavor competing metastable states and ease dynamical assembly. An iterative simulation-in-the-loop extension could identify kinetically relevant off-targets each round, supporting phase-specific control protocols that exploit, rather than fight, the system's kinetic path. A complementary direction is to represent discrete states probabilistically, as in expected Hamiltonian formulations~\cite{krueger2024generalized}, enabling optimization over distributions rather than assignments.

Taken together, these directions point toward a future in which complex self-assembling systems can be systematically engineered through a combination of physical theory, simulation, and gradient-based design.
By bridging accurate analytical models with scalable optimization, our framework moves the field closer to a general-purpose tool for programming matter at equilibrium.

\textit{Data availability.---}The simulation data needed to reproduce the results in this Letter are available in Ref.~\cite{guttieres2026data}, which also links to the source code repository.

\paragraph{Acknowledgments---}
This work was supported by the NSF AI Institute of Dynamic Systems (2112085), the Alfred P. Sloan Foundation under Grant No. G-2021-14198, and the Harvard MRSEC (NSF DMR-2011754). This research used resources provided by the Advanced Cyberinfrastructure Coordination Ecosystem: Services \& Support (ACCESS) program, supported by U.S. National Science Foundation Grants No. 2138259, No. 2138286, No. 2138307, No. 2137603, and No. 2138296, through allocation MTH250034 on the NCSA Delta system.

\section*{End Matter}

\textit{Differentiable yield computation.---}
We compute gradients of the yield with respect to $\bm{\theta}$ by decomposing the total derivative,
$\frac{dY}{d\bm{\theta}} = \frac{\partial Y}{\partial Z}\cdot\frac{dZ}{d\bm{\theta}}$.
We first consider $\frac{dZ}{d\bm{\theta}}$.
The cluster partition function $Z_s$ depends on $\bm{\theta}$ through its energy minimum and the vibrational spectrum (see Eq.~\ref{eq:part_fn}).
For the definitions of $\bm{\theta}$ considered in this work, the rotational entropy is independent of $\bm{\theta}$ and can be precomputed.
Therefore, we can directly compute $\frac{dZ}{d\bm{\theta}}$ via automatic differentiation, without differentiating the expensive sampling procedure necessary to compute $\widetilde{J}$.

For $\frac{\partial Y}{\partial Z}$, we leverage implicit differentiation through the fixed-point system defined by Eqs.~\ref{eq:selfconsistent2}--\ref{eq:selfconsistent1}, rather than directly differentiating through the unrolled numerical solver.
Because the yields $Y$ are the roots of a residual function $R(Y, Z) = 0$, the derivatives $dY/dZ$ can be computed by solving a linear system involving the Jacobian $(\partial R / \partial Y)^{-1}$, without back propagating through the iterative solver.
This circumvents the numerical instabilities introduced by differentiating iterative computations, and permits efficient gradient calculations using only the Jacobian of the residual function~\cite{blondel2021implicit}.

We implement this process for computing $dY/d\bm{\theta}$ in JAX~\cite{jax2018github}, an automatic differentiation framework.
We also solve the mapping from partition functions to equilibrium concentrations in log-space for stability.

\textit{Mass-action regularization for polymerization.---}
To penalize uncontrolled growth without explicit enumeration of all possible chains, we generalize classical equilibrium mass action theory~\cite{hagan2021selflimiting}.
For a single-monomer system, this theory describes the equilibrium concentration of $n$-mers as
\begin{equation}
c_n = n c_1^n e^{-n\beta \epsilon(n)} \label{eq:mass_original}
\end{equation}
where $c_1$ is the total building block monomer concentration and $\epsilon(n)$ is the mean free energy of each monomeric subunit within the chain such that $n\epsilon(n)$ is the free energy of the chain of length $n$.
Note that Equation \ref{eq:mass_original} is distinct from the mass-action constraint described by Equation \ref{eq:selfconsistent1}.
Equation \ref{eq:mass_original} can be extended to approximate the value of $c_n$ in the case of $M$ unique monomeric species, for which there are $M^n$ possible structures of length $n$ (up to symmetry).
This approximation is given by
\begin{equation}
c_{s} = n  e^{-\beta\epsilon(s)} \prod_{\alpha=1}^{M} c_\alpha^{N_{s, \alpha}} \label{eq:mass_act_true}
\end{equation}
where $c_s$ is the equilibrium concentration of structure $s$ (a candidate polymer of length $n$), $N_{s, \alpha}$ is the number of copies of monomer type $\alpha$ in structure $s$, and $c_\alpha$ is the equilibrium concentration of monomer type $\alpha$.
Since we perform gradient-based optimization, we can augment our objective function with an auxiliary loss term based on Equation \ref{eq:mass_act_true} to regularize the design problem (see SM).

\textit{Why the $n{=}4$ penalty suppresses longer chains.---}
Although formulated on the immediate overgrowth step ($n=4$), this penalty suppresses longer chains ($n\ge 4$) as observed in our simulations; we hypothesize this is because $n=4$ is a representative topological bottleneck for growth in this model. Maximizing trimer yield strengthens interactions that stabilize $n=3$, while the tetramer penalty weakens those same interactions that enable overgrowth. The resulting competition selects a narrow thermodynamic window: bonds favor the trimer but make incorporation of additional monomers thermodynamically unfavorable. Importantly, this argument does not depend on a specific kinetic pathway. Even if growth proceeds via cluster--cluster aggregation (e.g., $2{+}2$ or $2{+}3$), the merged aggregate must still realize the same local pairwise contacts and marginal free-energy gain associated with creating an $n\ge4$ state. Thus, making $n=4$ unfavorable typically keeps the marginal free-energy gain for further growth ($n=5,6,\ldots$) negative.

Without the $n{=}4$ penalty, the optimizer drives substantial, uncontrolled chain growth ($n\ge 4$) at intermediate temperatures (Fig.~\ref{fig:polymer_comparison}C, left). We suggest this reflects crossing a thermodynamic boundary: as temperature increases, the system exits a window of self-limiting stability and enters a regime of unbounded aggregation---where concentration and entropic gain favor elongation---before reaching the disassembly limit at high $T$. This mirrors the competition between finite structures and bulk phase separation in theoretical models of self-assembly~\cite{jacobs2016addressable, wilber2007patchy}. Our $n{=}4$ penalty (Fig.~\ref{fig:polymer_comparison}C, right) reshapes this landscape, widening the trimer stability window and preventing uncontrolled polymerization.

\nocite{kingma2014adam,goldbookSymmetry,wikipediaSymmetry,shaitan2022hidden,gilson2010symmetry,vandewiele2014symmetry,remiHOOMD2024,kruegerShellSim2023,optax2025,Martyna1994,Leimkuhler2015,Noe2019,frenkel2001,whitelam2015}

\newpage
\clearpage

\appendix
\onecolumngrid

\setcounter{figure}{0}
\renewcommand{\thefigure}{S\arabic{figure}}

\setcounter{table}{0}
\renewcommand{\thetable}{S\arabic{table}}

\setcounter{equation}{0}
\renewcommand{\theequation}{S\arabic{equation}}

\section*{Appendix A: Analytical Yield Calculation}

We employ a recently introduced~\cite{curatolo2023toolbox} analytical calculation to compute equilibrium assembly yields.
Here we provide additional details relating to our use of this calculation, i.e. the determination of symmetry numbers, the enumeration of off-target structures, and the mapping of partition functions to yields.

\subsection*{Symmetry Numbers $\sigma_s$}

The symmetry number $\sigma_s$ accounts for the number of rotationally indistinguishable configurations of a molecular assembly and serves as a correction factor in the partition function to avoid overcounting. 
In a system comprised of fully rigid bodies, it is defined as the number of distinct spatial arrangements that can be generated through rotation without yielding a distinguishable structure. 
This corresponds to the order of the molecule’s rotational symmetry group~\cite{goldbookSymmetry, wikipediaSymmetry, shaitan2022hidden, gilson2010symmetry}.
Below, we describe the determination of the symmetry numbers for each system considered in this work.

\paragraph*{Dimer System.} 
In Ref.~\cite{curatolo2023toolbox}, the authors explicitly construct the dimer system in a way that avoids zero modes.
This is achieved by placing three distinctly colored patches on one side of a core assembly consisting of three repulsive spheres arranged $120^{\circ}$ from each other.
This deliberate design ensures that no non-trivial rotation results in an indistinguishable configuration. 
As a result, the dimer cluster has a symmetry number $\sigma_s = 1$, simplifying the partition function and providing a clean baseline for yield prediction.

\paragraph*{Shell System.}

In our octahedral system, the target structure is a rigid shell assembled from monomers with directional patches. 
These shells approximate the polyhedral geometry of an octahedron, which corresponds to well-characterized point groups. 
In all of our case studies, monomers are treated as rigid and indistinguishable within an assembly, but no internal permutations or bond rearrangements are permitted.
Therefore, to quantify the symmetry number for a rigid cluster under these constraints, we adopt the formalism of Grimme et al.~\cite{vandewiele2014symmetry}, who define the total symmetry number for a cluster $s$ as:
\begin{equation}
\sigma_s = \left( \prod_i \sigma_{\text{int},i} \right) \cdot \sigma_{\text{ext}}.
\end{equation}
Here, $\sigma_{\text{int},i}$ accounts for the internal symmetry of monomer $i$, and $\sigma_{\text{ext}}$ reflects the external rotational symmetry of the overall structure.
The ideal octahedron belongs to the $O_h$ point group, which has an external symmetry number $\sigma_{\text{ext}} = 24$. These 24 operations include identity, 3-fold and 4-fold axis rotations, inversion, and improper rotations. For idealized shells with fully symmetric patch patterns, this symmetry number can be applied directly.
However, in our implementation, each monomer carries four patches divided into two  distinct species arranged in a 1–1–2–2 sequence, starting from one corner and proceeding clockwise around the core. This breaks the full internal symmetry of the monomer: rotating it about its center generally changes the identity of patch–patch interactions, even though the spatial geometry is preserved. 
Therefore, the monomers are not rotationally symmetric, so we assign an internal symmetry of $\sigma_{\text{int},i} = 1$ to each monomer.
As a result, we compute the total symmetry number $\sigma_s$ solely from the set of global rigid-body rotations that map the entire structure onto itself while preserving the identity of each patch. 
That is, we use:
\begin{equation}
\sigma_s = \sigma_{\text{ext}}, \quad \text{with} \quad \sigma_{\text{int},i} = 1 \ \forall i.
\end{equation}
To compute $\sigma_{\text{ext}}$ for each off-target shell structure, we implemented the following symmetry detection procedure:
\begin{enumerate}
    \item Load vertex positions and species identities.
    \item Re-center the positions per the structure's center of mass.
    \item Apply each of the 24 rotation matrices in the $O$ (octahedral) point group.
    \item For each rotation, check whether the rotated configuration is indistinguishable from the original by comparing the sorted coordinate sets within each species group.
\end{enumerate}
Only rotations that preserve both the spatial configuration and species assignment contribute to the symmetry number. 
This method allows us to compute the symmetry numbers even for partially symmetric or heterogeneous clusters. 
Although this approach remains an approximation, particularly for off-target clusters where deformation or partial bonding might lower effective symmetry, it provides a consistent and tractable estimate of $\sigma_s$ that respects both geometry and species identity.
Applying this methodology yields a symmetry number of 8 for the fully assembled shell and a symmetry number of 1 for all other intermediate cluster sizes.

\paragraph*{Polymerizing System.} 
Inspired by the polymerizing system in Ref. \cite{murugan2015undesired}, we define a system that exhibits non-self-limiting assembly based on an extension of the simple dimer system described above.
This system consists of monomers with similar symmetric patch arrangements: each monomer has three patches on each pole, with identical patch types on both sides. 
These are arranged at $120^{\circ}$ intervals around the attachment axis as in the dimer, resulting in a threefold internal rotational symmetry. 
To compute the symmetry number, we account for the following considerations:
\begin{itemize}
    \item Each monomer has threefold internal symmetry: $\sigma_{\text{int},i} = 3$.
    \item One monomer is treated as a reference and not counted toward internal redundancy.
    \item The chain as a whole admits 3 global rotations (e.g., around the chain axis), but these cancel out across all partition functions, and are factored out in our implementation.
\end{itemize}
Therefore, for a chain of $n$ monomers, the effective symmetry number is:
\begin{equation}
\label{eq:poly_sigma}
\sigma_s = 3^{n-1}.
\end{equation}
This correction accounts for the exponential increase in indistinguishable configurations due to repeated, internally symmetric monomers.

\subsection*{Off-Target Enumeration}

The analytical calculation requires the explicit enumeration of off-target assemblies.
Below, we describe the determination of these off-targets for each system.
In all cases, energy minimization is performed for each assembly to obtain the ground state.

\paragraph*{Dimer System.}
Delineating off-target structures in this case is trivial, since only two symmetric monomer types exist. 
These are enantiomeric -- mirror-related by patch arrangement -- and the sole target dimer corresponds to their correct attachment following the matching patch-color order.

\textit{Shell System.} Given the complete structure of the octahedral shell (see SM, Sec.B for a more detailed description of the geometry), we use a pruning procedure to enumerate off-target structures. 
Specifically, we consider the fully assembled shell (a rigid cluster of six monomers corresponding to the shell vertices) and generate connected subsets of this structure by recursively removing monomers while preserving connectivity. 
At each step, we remove one monomer and check whether the resulting subset remains a single connected component. 
If it does, the new configuration is considered a candidate off-target structure. 
This process is repeated until only a single monomer remains, resulting in a hierarchy of fully connected off-target structures of sizes $1-5$.

To make the calculation tractable, we make a simplifying approximation: for each cluster size  $n$ , we retain only one representative connected configuration. 
While in principle there may be multiple geometrically distinct off-targets of the same size (e.g., several ways to select 4 connected vertices from the full shell), we assume that a single representative configuration sufficiently captures the contribution to the partition function for that size class. 
This is likely owing to the large space of candidate off-targets that are not included in the theoretical calculation. Because the theoretical model employs a truncated state space, the optimizer drives the predicted yield to $\approx 100\%$ within this simplified landscape. While effective for identifying optimal parameters, this theoretical prediction systematically overestimates the yield by neglecting un-enumerated off-targets. We therefore omit these theoretical curves in Figures 3C and 4C, as they would appear as uninformative flat lines relative to the realistic, competition-limited MD results.

\paragraph*{Polymerizing System.}

To build the list of polymerized clusters included in the optimization, we use a combinatorial enumeration procedure that generates all valid connected sequences of monomers up to a specified maximum size. Although the polymer monomers share the same physical structure as the dimer monomers, with the only difference being that both sides of the polymer monomer carry a tri-patch site, for simplicity our enumeration algorithm represents each monomer using a central vertex label and two patch indices only. 
These indices specify how the monomer connects to its neighbors in the chain. 
The main steps of the algorithm are as follows:
\begin{itemize}
    \item \textit{Monomer Representation.} We define $N$ monomer types (e.g., $\alpha$, $\beta$, $\gamma$), each with a forward and reverse orientation (e.g., $\alpha^\prime$, $\beta^\prime$, $\gamma^\prime$). The forward version of monomer $X$ is denoted $X$, and the reverse (flipped) version is denoted $X'$; their patch indices are reversed accordingly. This effectively doubles the set of monomer building blocks and allows the algorithm to account for orientational degrees of freedom.
    \item \textit{Sequence Enumeration.} We construct all ordered sequences of monomers of size $1 \leq n \leq n_{\max}$ using Cartesian products of the full monomer set (forward + reverse). Each sequence is treated as a candidate cluster.
    \item \textit{Symmetry Pruning.} To avoid double-counting symmetric structures, we discard any sequence that is a mirror image of one already included. 
    Mirror images are defined as the reverse of the monomer sequence with all monomer orientations flipped (i.e., $X \leftrightarrow X'$).
    \item \textit{Species Encoding.} Each valid cluster is converted into a numeric representation based on its monomer patch sequence. These are stored as the species identifiers used throughout the optimization pipeline.
    \item \textit{Symmetry Number Assignment.} For the polymer system, the symmetry number $\sigma_s$ of a cluster of size $n$ is computed using Eq.~\ref{eq:poly_sigma} reflecting the threefold internal symmetry of each monomer beyond a fixed reference unit. 
\end{itemize}
We explicitly enumerate all chains up to length $n=3$, resulting in $132$ total clusters included in the analytical calculation.
The number of possible structures grows exponentially with chain length, e.g. there are $666$ structures of length $n=4$.
We circumvent the costly partition function calculation for larger chains by applying a mass action penalty for these $n=4$ structures (see SM, Sec.D).
Note that this mass action penalty still requires enumerating the possible structures of a length $n=4$.

\subsection*{Yield Calculation}

The partition function $Z_s$ describes the statistical weight of an individual assembly $s$, however actual self-assembly processes feature many clusters forming simultaneously and competing for the same pool of building blocks.
In this work, we define the equilibrium yield of cluster $s$, $Y_s$, as the likelihood of sampling $s$ upon randomly sampling a cluster in equilibrium.

Following the derivation in ~\cite{curatolo2023toolbox} we define the equilibrium yield of a particular structure in the grand canonical ensemble as follows:
\begin{align}
Y_s=\frac{\left(\prod_{\alpha}\tilde{c}_\alpha^{N_{s,\alpha}}\right)Z_s}{\mathcal{Q}}
\end{align}
where $\tilde{c}_{\alpha}$ is the total concentration of monomer $\alpha$ in the system, ${N_{s,\alpha}}$ is the number of monomers type $\alpha$ in structure $s$, and $\mathcal{Q}$ is the grand partition function. 
Given this definition, we map the structure partition functions to equilibrium concentrations by numerically solving the system of equations derived by Curatolo et al.~\cite{curatolo2023toolbox} and described in the main text (Equations 4 and 5). 
We solve this system of equations using either the \texttt{GradientDescent} (dimer and polymerizing systems) or \texttt{LBFGS} (shell system) solvers in the Python \texttt{jaxopt} library, terminating when the relative change in all cluster concentrations $c_s$ falls below $10^{-6}$.
We perform this procedure in log-space for numerical stability.

The baseline cost function for this procedure is the L2-norm of residuals for the system of equations.
To promote uniform convergence across species, we also augment the cost function with a term describing the variance in the residuals across species.
Specifically, for the dimer and polymerizing systems, the total cost function is 
$
R_{\text{tot}} = \|\mathbf{r}\|_2 + \mathrm{Var}(\mathbf{r}),
$
where $\mathbf{r}$ denotes the residuals. 
For the shell system, the same formulation is applied but with an additional weighting on the monomeric term and a stronger variance regularization, i.e., 
$
R_{\text{tot}} = \|w \odot \mathbf{r}\|_2 + 50\,\mathrm{Var}(\mathbf{r}),
$
where $w = [10, 1, 1, 1, 1]$. 
This weighting reflects the fact that monomers are the most probable non-assembled configuration \cite{hagan2021selflimiting} and thus dominate the equilibrium landscape. 

\clearpage
\newpage

\section*{Appendix B: System Descriptions}

Each of our model systems consists of rigid ``patchy'' particles. 
In all cases, the main body of a monomer is composed of central vertex particles that interact exclusively according to same-type repulsion. 
The patches interact with one another via an attractive Morse potential. 
Below, we describe the geometry and interaction potentials for the three model systems used in this study.

\textit{Dimer System.}
This system is adapted directly from the model presented in Ref.~\cite{curatolo2023toolbox}. There exist only two monomer species in the system. Each monomer is made of a main repulsive body of three vertex particles with three distinct patches arranged exclusively on one of the ``faces'' of the main body. Patches only attract to others of the same color (self-specific binding), and the second monomer species in the system is the mirror image of the first. This specific three fold geometry is chosen such that no incomplete attraction can possibly occur. 
This toy system is deliberately simple: if monomers attract, they engage all patches simultaneously, forming only the target dimer. 
More specifically, patch interactions are described by a Morse potential:
\begin{equation}
U_{\text{Morse}}(r)
= D_0\!\left[\left(1-e^{-\alpha\,(r-r_0)}\right)^2 - 1\right]
= D_0\!\left(e^{-2\alpha (r-r_0)} - 2 e^{-\alpha (r-r_0)}\right),
\end{equation}
where $D_0 = \varepsilon_{ij}$ are pair-specific well depths, and $r_0 = 0$, $\alpha = 5.0$ are shared shape parameters. 
Core-core interactions are described by a short-range, soft-sphere potential:
\begin{equation}
U_{\text{rep}}(r) =
\begin{cases}
\dfrac{A}{\alpha\, r_{\mathrm{cut}}}\,(r_{\mathrm{max}} - r)^{\alpha}\,S(r), & r < r_{\mathrm{max}}, \\[6pt]
0, & r \ge r_{\mathrm{max}}.
\end{cases}
\end{equation}
where $S(r)$ is a smoothing function that brings the potential continuously to zero at the cutoff,
\begin{equation}
S(r) = \frac{1}{1 + \exp[-\kappa((r - r_{\min})/(r_{\max}-r_{\min}) - 0.5)]},
\end{equation}
with $\kappa = 10$ controlling the steepness of the transition. The parameters are $A = 500.0$, $\alpha = 2.5$, $r_{\min} = 0.0$, $r_{\max} = 2.0$, and $r_{\mathrm{cut}} = 6.0$. Both the Morse and repulsive potentials are shifted such that $U(r_{\mathrm{cut}}) = 0$.

\textit{Shell System.} 
The octahedral system is directly adapted from the octahedral construction described in  Ref. \cite{kruegerShellSim2023}. 
We modified the initial geometry to include two distinct patch types which only allow same-type binding. 
Therefore, each monomer consists of one central vertex and four directional patches arranged tetrahedrally, in a 1-1-2-2 pattern around the core. 
This change both strengthens correct assembly and increases the system's combinatorial complexity. 
The energy function combines soft-sphere repulsion between rigid-body centers with species-specific Morse attractions between patches.
Soft-sphere repulsion between core particles follows:
\begin{align}
U_{\text{soft}}(r) = \epsilon_{\text{soft}} \left( \frac{\sigma}{r} \right)^{12},
  \quad \epsilon_{\text{soft}} = 10^4, \quad \sigma = 2.0  
\end{align}
while Morse attraction between same-type patches is described as:
\begin{align} U_{\text{Morse}}(r) = \epsilon_{\text{Morse}}
  \left( e^{-2\alpha (r - r_0)} - 2 e^{-\alpha (r - r_0)} \right),
  \quad \epsilon_{\text{Morse}} = 10.0, \quad \alpha = 2.0, \quad r_{\text{cut}} = 12.0,
\end{align}

\textit{Polymerizing System}
Inspired by the polymerizing system in Ref. \cite{murugan2015undesired}, the polymerizing case extends the dimer model by placing patches on opposite sides of the central core, with patches on the same side sharing the same color, and allowing all patch species to interact attractively with one another.
We define three distinct monomer species, each bearing two distinct-colored patches, resulting in a total of seven patch species including the vertex type which makes up the main body.
The interaction potentials and parameters are identical to those used in the dimer case (Morse attraction and short-range repulsion). However, because patches are permitted to bind irrespective of color, the patch interaction matrix expands to a symmetric $6\times6$ matrix.

\clearpage
\newpage

\section*{Appendix C: Optimization Details}

For all optimizations, we perform gradient descent using a system-specific objective function.
Gradients with respect to the control parameters are automatically computed via JAX~\cite{jax2018github}.
Rather than computing derivatives through the unrolled numerical procedure described in Appendix B for mapping partition functions to concentrations, we apply implicit differentiation~\cite{blondel2021implicit}.
Rather than defining the optimality condition for implicit differentiation as the gradient of the cost function, which is standard for root finding, we define the optimality condition as the residuals themselves.
This serves as a stronger and therefore higher-signal optimality condition, and is valid as the system of equations is fully determined by the partition functions.
We implement implicit differentiation via the  \texttt{jaxopt.implicit\_diff.custom\_root} primitive. 

\begin{table*}[t]
\caption{\textbf{Optimized parameters for Dimer system -- Fixed Temperature.} Interaction strength parameters ($\epsilon$) optimized to achieve target yields at fixed temperatures. All three patch interactions (red, blue, green) are optimized independently but converge to similar values.}
\label{tab:params_dimer_fixedkt}
\begin{ruledtabular}
\begin{tabular}{c c c c c c}
\textbf{Target} & \textbf{Predicted} & $\bm{\epsilon_{\text{black}}}$ & $\bm{\epsilon_{\text{blue}}}$ & $\bm{\epsilon_{\text{green}}}$ & \textbf{$kT$} \\
\textbf{Yield} & \textbf{Yield} & & & & \\
\hline
\multicolumn{6}{c}{\textit{$kT = 0.5$}} \\
0.10 & 0.108 & 3.62 & 3.62 & 3.62 & 0.5 \\
0.20 & 0.195 & 3.80 & 3.80 & 3.80 & 0.5 \\
0.30 & 0.300 & 3.95 & 3.95 & 3.95 & 0.5 \\
0.40 & 0.398 & 4.08 & 4.08 & 4.08 & 0.5 \\
0.50 & 0.501 & 4.19 & 4.19 & 4.19 & 0.5 \\
0.60 & 0.601 & 4.34 & 4.34 & 4.34 & 0.5 \\
0.70 & 0.703 & 4.48 & 4.48 & 4.48 & 0.5 \\
0.80 & 0.796 & 4.66 & 4.66 & 4.66 & 0.5 \\
0.90 & 0.899 & 4.96 & 4.96 & 4.96 & 0.5 \\
1.00 & 0.993 & 5.97 & 5.97 & 5.97 & 0.5 \\
\hline
\multicolumn{6}{c}{\textit{$kT = 1.0$}} \\
0.10 & 0.100 & 7.200 & 7.200 & 7.200 & 1.0 \\
0.20 & 0.198 & 7.608 & 7.609 & 7.609 & 1.0 \\
0.30 & 0.302 & 7.908 & 7.909 & 7.909 & 1.0 \\
0.40 & 0.397 & 8.151 & 8.151 & 8.151 & 1.0 \\
0.50 & 0.499 & 8.381 & 8.382 & 8.382 & 1.0 \\
0.60 & 0.594 & 8.677 & 8.678 & 8.678 & 1.0 \\
0.70 & 0.699 & 8.951 & 8.950 & 8.950 & 1.0 \\
0.80 & 0.794 & 9.289 & 9.289 & 9.289 & 1.0 \\
0.90 & 0.860 & 9.622 & 9.622 & 9.622 & 1.0 \\
1.00 & 0.928 & 10.148 & 10.147 & 10.147 & 1.0 \\
\hline
\multicolumn{6}{c}{\textit{$kT = 1.5$}} \\
0.10 & 0.098 & 10.84 & 10.84 & 10.84 & 1.5 \\
0.20 & 0.199 & 11.40 & 11.41 & 11.41 & 1.5 \\
0.30 & 0.296 & 11.86 & 11.86 & 11.86 & 1.5 \\
0.40 & 0.397 & 12.24 & 12.24 & 12.24 & 1.5 \\
0.50 & 0.500 & 12.61 & 12.61 & 12.61 & 1.5 \\
0.60 & 0.594 & 12.98 & 12.98 & 12.98 & 1.5 \\
0.70 & 0.668 & 13.28 & 13.28 & 13.28 & 1.5 \\
0.80 & 0.791 & 13.95 & 13.95 & 13.95 & 1.5 \\
0.80 & 0.762 & 13.75 & 13.75 & 13.75 & 1.5 \\
0.90 & 0.898 & 14.81 & 14.81 & 14.81 & 1.5 \\
\end{tabular}
\end{ruledtabular}
\end{table*}

\begin{table*}[t]
\caption{\textbf{Optimized parameters for Dimer system -- Fixed Interaction Strength.} Temperature ($kT$) optimized to achieve target yields at fixed interaction strengths ($\epsilon$). Demonstrates temperature-based yield control.}
\label{tab:params_dimer_fixedeps}
\begin{ruledtabular}
\begin{tabular}{c c c c}
\textbf{Target Yield} & \textbf{Predicted Yield} & \textbf{Optimized $kT$} & $\bm{\epsilon}$ \textbf{(fixed)} \\
\hline
\multicolumn{4}{c}{\textit{$\epsilon = 3.0$}} \\
0.10 & 0.101 & 0.415 & 3.0 \\
0.20 & 0.193 & 0.395 & 3.0 \\
0.30 & 0.300 & 0.379 & 3.0 \\
0.40 & 0.397 & 0.368 & 3.0 \\
0.50 & 0.501 & 0.356 & 3.0 \\
0.60 & 0.600 & 0.346 & 3.0 \\
0.70 & 0.700 & 0.334 & 3.0 \\
0.80 & 0.800 & 0.321 & 3.0 \\
0.90 & 0.900 & 0.302 & 3.0 \\
1.00 & 1.000 & 0.164 & 3.0 \\
\hline
\multicolumn{4}{c}{\textit{$\epsilon = 6.5$}} \\
0.10 & 0.100 & 0.900 & 6.5 \\
0.20 & 0.201 & 0.853 & 6.5 \\
0.30 & 0.300 & 0.823 & 6.5 \\
0.40 & 0.401 & 0.797 & 6.5 \\
0.50 & 0.500 & 0.774 & 6.5 \\
0.60 & 0.601 & 0.750 & 6.5 \\
0.70 & 0.701 & 0.725 & 6.5 \\
0.80 & 0.800 & 0.696 & 6.5 \\
0.90 & 0.900 & 0.655 & 6.5 \\
1.00 & 0.994 & 0.539 & 6.5 \\
\hline
\multicolumn{4}{c}{\textit{$\epsilon = 10.0$}} \\
0.10 & 0.100 & 1.387 & 10.0 \\
0.20 & 0.200 & 1.315 & 10.0 \\
0.30 & 0.300 & 1.267 & 10.0 \\
0.40 & 0.401 & 1.227 & 10.0 \\
0.50 & 0.501 & 1.191 & 10.0 \\
0.60 & 0.600 & 1.155 & 10.0 \\
0.70 & 0.700 & 1.117 & 10.0 \\
0.80 & 0.799 & 1.072 & 10.0 \\
0.90 & 0.900 & 1.009 & 10.0 \\
1.00 & 0.992 & 0.849 & 10.0 \\
\end{tabular}
\end{ruledtabular}
\end{table*}

\begin{table*}[t]
\caption{\textbf{Optimized parameters for Shell system -- Temperature-Controlled Assembly/Disassembly.} Parameters optimized for switchlike behavior: maximizing assembly at $kT_{\text{low}}$ while minimizing assembly at $kT_{\text{high}}$. Both patch types (green and yellow) share identical interaction strengths.}
\label{tab:params_shell}
\begin{ruledtabular}
\begin{tabular}{c c c c c c}
$\bm{kT_{\textbf{low}}}$ & \textbf{\% Increase} & $\bm{kT_{\textbf{high}}}$ & $\bm{\epsilon}$ & \textbf{Predicted Yield} & \textbf{Predicted Yield} \\
& & & & \textbf{at $kT_{\text{high}}$ (\%)} & \textbf{at $kT_{\text{low}}$ (\%)} \\
\hline
\multicolumn{6}{c}{\textit{Baseline $kT_{\text{low}} = 1.0$}} \\
1.0 & 5 & 1.05 & 7.70 & 5.09 & 99.47 \\
1.0 & 10 & 1.10 & 8.58 & 5.60 & 99.65 \\
1.0 & 15 & 1.15 & 9.26 & 5.05 & 100.00 \\
1.0 & 20 & 1.20 & 9.01 & 0.015 & 99.93 \\
1.0 & 25 & 1.25 & 8.74 & 0.006 & 99.22 \\
1.0 & 30 & 1.30 & 8.77 & 0.003 & 99.99 \\
1.0 & 35 & 1.35 & 9.02 & $1.6 \times 10^{-5}$ & 99.91 \\
1.0 & 40 & 1.40 & 8.62 & $1.0 \times 10^{-5}$ & 99.98 \\
1.0 & 45 & 1.45 & 8.90 & $9.5 \times 10^{-7}$ & 99.98 \\
1.0 & 50 & 1.50 & 8.72 & $2.5 \times 10^{-6}$ & 99.99 \\
\hline
\multicolumn{6}{c}{\textit{Baseline $kT_{\text{low}} = 1.25$}} \\
1.25 & 5 & 1.31 & 12.24 & 5.52 & 99.77 \\
1.25 & 10 & 1.38 & 13.96 & 5.13 & 99.78 \\
1.25 & 15 & 1.44 & 14.70 & 4.87 & 99.99 \\
1.25 & 20 & 1.50 & 12.55 & 0.0003 & 100.00 \\
1.25 & 25 & 1.56 & 13.88 & $2.2 \times 10^{-7}$ & 99.02 \\
1.25 & 30 & 1.63 & 12.65 & 4.96 & 99.99 \\
1.25 & 35 & 1.69 & 14.28 & 4.33 & 100.00 \\
1.25 & 40 & 1.75 & 13.79 & 0.006 & 99.94 \\
1.25 & 45 & 1.81 & 13.01 & $5.9 \times 10^{-5}$ & 99.99 \\
1.25 & 50 & 1.88 & 12.98 & $1.0 \times 10^{-6}$ & 99.99 \\
\hline
\multicolumn{6}{c}{\textit{Baseline $kT_{\text{low}} = 1.5$}} \\
1.5 & 5 & 1.58 & 15.40 & 4.75 & 99.97 \\
1.5 & 10 & 1.65 & 17.25 & 5.15 & 99.98 \\
1.5 & 15 & 1.73 & 18.05 & 3.78 & 99.99 \\
1.5 & 20 & 1.80 & 17.89 & 4.37 & 100.00 \\
1.5 & 25 & 1.88 & 18.83 & 4.30 & 99.99 \\
1.5 & 30 & 1.95 & 18.25 & $6.1 \times 10^{-6}$ & 94.00 \\
1.5 & 35 & 2.03 & 18.10 & $3.8 \times 10^{-5}$ & 100.00 \\
1.5 & 40 & 2.10 & 18.09 & $4.4 \times 10^{-6}$ & 99.98 \\
1.5 & 45 & 2.18 & 16.59 & $1.2 \times 10^{-8}$ & 99.80 \\
1.5 & 50 & 2.25 & 17.41 & $3.6 \times 10^{-8}$ & 99.91 \\
\end{tabular}
\end{ruledtabular}
\end{table*}

\begin{table*}[t]
\footnotesize
\caption{\textbf{Complete optimized parameters for Polymer system -- With Mass Action Regularization.} All 21 pairwise interaction strengths and monomer concentrations for the three-species ($\alpha$, $\beta$, $\gamma$) system. Patch numbering: 1,2 ($\alpha$); 3,4 ($\beta$); 5,6 ($\gamma$). Target bonds (bold rows): $\epsilon_{2,3}$ ($\alpha_2 \to \beta_1$) and $\epsilon_{4,5}$ ($\beta_2 \to \gamma_1$).}
\label{tab:params_polymer_mass}
\begin{ruledtabular}
\begin{tabular}{l c c c c c c c c c c c}
\textbf{Parameter} & \textbf{0.50} & \textbf{0.55} & \textbf{0.60} & \textbf{0.65} & \textbf{0.70} & \textbf{0.75} & \textbf{0.80} & \textbf{0.85} & \textbf{0.90} & \textbf{0.95} & \textbf{1.00} \\
 & \multicolumn{11}{c}{$kT$ (reduced temperature)} \\
\hline
\multicolumn{12}{l}{\textit{Optimized Yield}} \\
Yield & 1.000 & 0.992 & 1.000 & 0.999 & 1.000 & 0.911 & 0.707 & 0.589 & 0.697 & 1.000 & 0.700 \\
\hline
\multicolumn{12}{l}{\textit{Target Bonds}} \\
$\bm{\epsilon_{2,3}}$ & \textbf{7.551} & \textbf{5.453} & \textbf{7.058} & \textbf{7.847} & \textbf{8.331} & \textbf{6.395} & \textbf{6.299} & \textbf{6.565} & \textbf{7.239} & \textbf{7.558} & \textbf{8.234} \\
$\bm{\epsilon_{4,5}}$ & \textbf{7.552} & \textbf{5.453} & \textbf{6.909} & \textbf{7.754} & \textbf{8.514} & \textbf{6.395} & \textbf{6.299} & \textbf{6.565} & \textbf{7.239} & \textbf{7.558} & \textbf{8.234} \\
\hline
\multicolumn{12}{l}{\textit{Cross-Species Interactions}} \\
$\epsilon_{1,2}$ & 0.797 & 0.376 & 1.444 & 2.786 & 2.610 & 2.198 & 2.100 & 2.397 & 2.128 & 1.288 & 0.269 \\
$\epsilon_{1,3}$ & 0.795 & 1.368 & 1.321 & 2.479 & 1.783 & 2.198 & 2.100 & 2.397 & 2.318 & 1.425 & 0.317 \\
$\epsilon_{1,4}$ & 0.795 & 1.368 & 1.321 & 2.478 & 1.783 & 2.198 & 2.100 & 2.397 & 2.318 & 1.425 & 0.311 \\
$\epsilon_{1,5}$ & 0.797 & 0.376 & 1.852 & 2.895 & 3.424 & 2.198 & 2.100 & 2.397 & 2.128 & 1.289 & 0.276 \\
$\epsilon_{1,6}$ & 0.797 & 0.376 & 1.852 & 2.895 & 3.424 & 2.198 & 2.100 & 2.397 & 2.128 & 1.289 & 0.272 \\
$\epsilon_{2,4}$ & 0.796 & 1.369 & 1.321 & 2.478 & 1.783 & 2.198 & 2.100 & 2.397 & 2.318 & 1.425 & 0.307 \\
$\epsilon_{2,5}$ & 0.797 & 0.376 & 1.852 & 2.895 & 3.424 & 2.198 & 2.100 & 2.397 & 2.128 & 1.289 & 0.272 \\
$\epsilon_{2,6}$ & 0.797 & 0.377 & 1.852 & 2.895 & 3.424 & 2.198 & 2.100 & 2.397 & 2.128 & 1.288 & 0.269 \\
$\epsilon_{3,4}$ & 0.795 & 1.941 & 1.835 & 0.966 & 1.105 & 2.198 & 2.100 & 2.396 & 2.467 & 1.554 & 0.543 \\
$\epsilon_{3,5}$ & 0.795 & 1.378 & 1.587 & 1.277 & 2.563 & 2.198 & 2.100 & 2.397 & 2.318 & 1.425 & 0.323 \\
$\epsilon_{3,6}$ & 0.795 & 1.378 & 1.587 & 1.276 & 2.563 & 2.198 & 2.100 & 2.397 & 2.318 & 1.425 & 0.317 \\
$\epsilon_{4,6}$ & 0.795 & 1.378 & 1.587 & 1.277 & 2.563 & 2.198 & 2.100 & 2.397 & 2.318 & 1.425 & 0.313 \\
$\epsilon_{5,6}$ & 0.797 & 0.377 & 1.741 & 1.600 & 3.317 & 2.198 & 2.100 & 2.397 & 2.128 & 1.289 & 0.276 \\
\hline
\multicolumn{12}{l}{\textit{Self Interactions}} \\
$\epsilon_{1,1}$ & 0.797 & 0.372 & 1.445 & 2.784 & 2.611 & 2.198 & 2.100 & 2.397 & 2.125 & 1.286 & 0.271 \\
$\epsilon_{2,2}$ & 0.797 & 0.373 & 1.445 & 2.785 & 2.611 & 2.198 & 2.100 & 2.397 & 2.125 & 1.286 & 0.265 \\
$\epsilon_{3,3}$ & 0.795 & 1.947 & 1.845 & 0.967 & 1.092 & 2.198 & 2.100 & 2.396 & 2.471 & 1.558 & 0.567 \\
$\epsilon_{4,4}$ & 0.795 & 1.947 & 1.845 & 0.968 & 1.092 & 2.198 & 2.100 & 2.396 & 2.471 & 1.558 & 0.548 \\
$\epsilon_{5,5}$ & 0.797 & 0.374 & 1.742 & 1.601 & 3.316 & 2.198 & 2.100 & 2.397 & 2.125 & 1.287 & 0.279 \\
$\epsilon_{6,6}$ & 0.798 & 0.374 & 1.742 & 1.601 & 3.316 & 2.198 & 2.100 & 2.397 & 2.125 & 1.286 & 0.272 \\
\hline
\multicolumn{12}{l}{\textit{Monomer Concentrations ($\times 10^{-5}$)}} \\
$c_{\alpha}$ & 3.33 & 3.33 & 3.02 & 6.80 & 2.67 & 3.33 & 3.33 & 3.33 & 3.32 & 3.32 & 3.32 \\
$c_{\beta}$ & 3.33 & 3.33 & 3.44 & 1.47 & 2.35 & 3.34 & 3.33 & 3.34 & 3.36 & 3.35 & 3.36 \\
$c_{\gamma}$ & 3.33 & 3.33 & 3.55 & 1.73 & 4.98 & 3.33 & 3.33 & 3.33 & 3.32 & 3.32 & 3.32 \\
\end{tabular}
\end{ruledtabular}
\vspace{0.1cm}
\textit{Note:} At $kT=0.65$--$0.70$, non-stoichiometric concentrations emerge. At $kT=0.75$--$0.85$, all interactions converge to nearly identical values.
\end{table*}

\begin{table*}[t]
\footnotesize
\caption{\textbf{Complete optimized parameters for Polymer system -- Without Mass Action Regularization.} Same system as Table S4 but optimized without mass action penalty. Note degraded yield at $kT=0.95$ (0.537 vs. 1.000).}
\label{tab:params_polymer_nomass}
\begin{ruledtabular}
\begin{tabular}{l c c c c c c c c c c c}
\textbf{Parameter} & \textbf{0.50} & \textbf{0.55} & \textbf{0.60} & \textbf{0.65} & \textbf{0.70} & \textbf{0.75} & \textbf{0.80} & \textbf{0.85} & \textbf{0.90} & \textbf{0.95} & \textbf{1.00} \\
 & \multicolumn{11}{c}{$kT$ (reduced temperature)} \\
\hline
\multicolumn{12}{l}{\textit{Optimized Yield}} \\
Yield & 1.000 & 0.998 & 1.000 & 0.999 & 1.000 & 0.911 & 0.707 & 0.589 & 0.697 & \textbf{0.537} & 0.700 \\
\hline
\multicolumn{12}{l}{\textit{Target Bonds}} \\
$\bm{\epsilon_{2,3}}$ & \textbf{7.551} & \textbf{6.336} & \textbf{7.186} & \textbf{7.847} & \textbf{8.331} & \textbf{6.395} & \textbf{6.299} & \textbf{6.565} & \textbf{7.239} & \textbf{7.429} & \textbf{8.234} \\
$\bm{\epsilon_{4,5}}$ & \textbf{7.552} & \textbf{6.347} & \textbf{6.941} & \textbf{7.754} & \textbf{8.514} & \textbf{6.395} & \textbf{6.299} & \textbf{6.565} & \textbf{7.239} & \textbf{7.429} & \textbf{8.234} \\
\hline
\multicolumn{12}{l}{\textit{Cross-Species Interactions}} \\
$\epsilon_{1,2}$ & 0.797 & 1.142 & 0.885 & 2.786 & 2.610 & 2.198 & 2.100 & 2.397 & 2.128 & 1.188 & 0.268 \\
$\epsilon_{1,3}$ & 0.795 & 1.273 & 0.924 & 2.479 & 1.783 & 2.198 & 2.100 & 2.397 & 2.318 & 1.282 & 0.315 \\
$\epsilon_{1,4}$ & 0.795 & 1.273 & 0.923 & 2.478 & 1.783 & 2.198 & 2.100 & 2.397 & 2.318 & 1.282 & 0.310 \\
$\epsilon_{1,5}$ & 0.797 & 1.150 & 2.485 & 2.895 & 3.424 & 2.198 & 2.100 & 2.397 & 2.128 & 1.188 & 0.276 \\
$\epsilon_{1,6}$ & 0.797 & 1.150 & 2.485 & 2.895 & 3.424 & 2.198 & 2.100 & 2.397 & 2.128 & 1.188 & 0.272 \\
$\epsilon_{2,4}$ & 0.796 & 1.273 & 0.923 & 2.478 & 1.783 & 2.198 & 2.100 & 2.397 & 2.318 & 1.282 & 0.306 \\
$\epsilon_{2,5}$ & 0.797 & 1.150 & 2.485 & 2.895 & 3.424 & 2.198 & 2.100 & 2.397 & 2.128 & 1.188 & 0.272 \\
$\epsilon_{2,6}$ & 0.797 & 1.150 & 2.485 & 2.895 & 3.424 & 2.198 & 2.100 & 2.397 & 2.128 & 1.188 & 0.269 \\
$\epsilon_{3,4}$ & 0.795 & 1.348 & 1.974 & 0.966 & 1.105 & 2.198 & 2.100 & 2.396 & 2.467 & 1.380 & 0.543 \\
$\epsilon_{3,5}$ & 0.795 & 1.280 & 2.535 & 1.277 & 2.563 & 2.198 & 2.100 & 2.397 & 2.318 & 1.282 & 0.323 \\
$\epsilon_{3,6}$ & 0.795 & 1.280 & 2.535 & 1.276 & 2.563 & 2.198 & 2.100 & 2.397 & 2.318 & 1.282 & 0.319 \\
$\epsilon_{4,6}$ & 0.795 & 1.280 & 2.538 & 1.277 & 2.563 & 2.198 & 2.100 & 2.397 & 2.318 & 1.282 & 0.313 \\
$\epsilon_{5,6}$ & 0.797 & 1.158 & 2.510 & 1.600 & 3.317 & 2.198 & 2.100 & 2.397 & 2.128 & 1.188 & 0.276 \\
\hline
\multicolumn{12}{l}{\textit{Self Interactions}} \\
$\epsilon_{1,1}$ & 0.797 & 1.139 & 0.881 & 2.784 & 2.611 & 2.198 & 2.100 & 2.397 & 2.125 & 1.187 & 0.272 \\
$\epsilon_{2,2}$ & 0.797 & 1.139 & 0.881 & 2.785 & 2.611 & 2.198 & 2.100 & 2.397 & 2.125 & 1.187 & 0.265 \\
$\epsilon_{3,3}$ & 0.795 & 1.349 & 1.977 & 0.967 & 1.092 & 2.198 & 2.100 & 2.396 & 2.471 & 1.383 & 0.574 \\
$\epsilon_{4,4}$ & 0.795 & 1.349 & 1.977 & 0.968 & 1.092 & 2.198 & 2.100 & 2.396 & 2.471 & 1.383 & 0.545 \\
$\epsilon_{5,5}$ & 0.797 & 1.155 & 2.509 & 1.601 & 3.316 & 2.198 & 2.100 & 2.397 & 2.125 & 1.187 & 0.279 \\
$\epsilon_{6,6}$ & 0.798 & 1.155 & 2.508 & 1.601 & 3.316 & 2.198 & 2.100 & 2.397 & 2.125 & 1.187 & 0.272 \\
\hline
\multicolumn{12}{l}{\textit{Monomer Concentrations ($\times 10^{-5}$)}} \\
$c_{\alpha}$ & 3.33 & 3.33 & 1.34 & 6.80 & 2.67 & 3.33 & 3.33 & 3.33 & 3.32 & 3.33 & 3.32 \\
$c_{\beta}$ & 3.33 & 3.33 & 6.85 & 1.47 & 2.35 & 3.34 & 3.33 & 3.34 & 3.36 & 3.35 & 3.36 \\
$c_{\gamma}$ & 3.33 & 3.33 & 1.81 & 1.73 & 4.98 & 3.33 & 3.33 & 3.33 & 3.32 & 3.33 & 3.32 \\
\end{tabular}
\end{ruledtabular}
\vspace{0.1cm}
\textit{Note:} Without mass action regularization, yield drops to 0.537 at $kT=0.95$ (vs. 1.000 in Table S4), demonstrating the importance of the penalty term.
\end{table*}

For all optimizations, we use an Adam optimizer~\cite{optax2025, kingma2014adam} with a learning rate of $10^{-2}$ to $8 \times 10^{-2}$ depending on the system sensitivity and chosen initial parameters.
Below, we describe the system-specific objective functions and hyperparameters:

\paragraph*{Dimer System.}
The loss is defined as the absolute difference between the predicted and desired yield:
\begin{align}
\mathcal{L}_{\text{dimer}} = \left|Y_{\text{target}} - Y_{\text{desired}}\right|.
\end{align}
Convergence is reliably achieved by approximately 300 iterations (see Fig. 2b), with all predicted yields falling within $1\%$ of the target yield value.

\begin{figure}[b]
    \centering
    \includegraphics[width=0.45\textwidth]{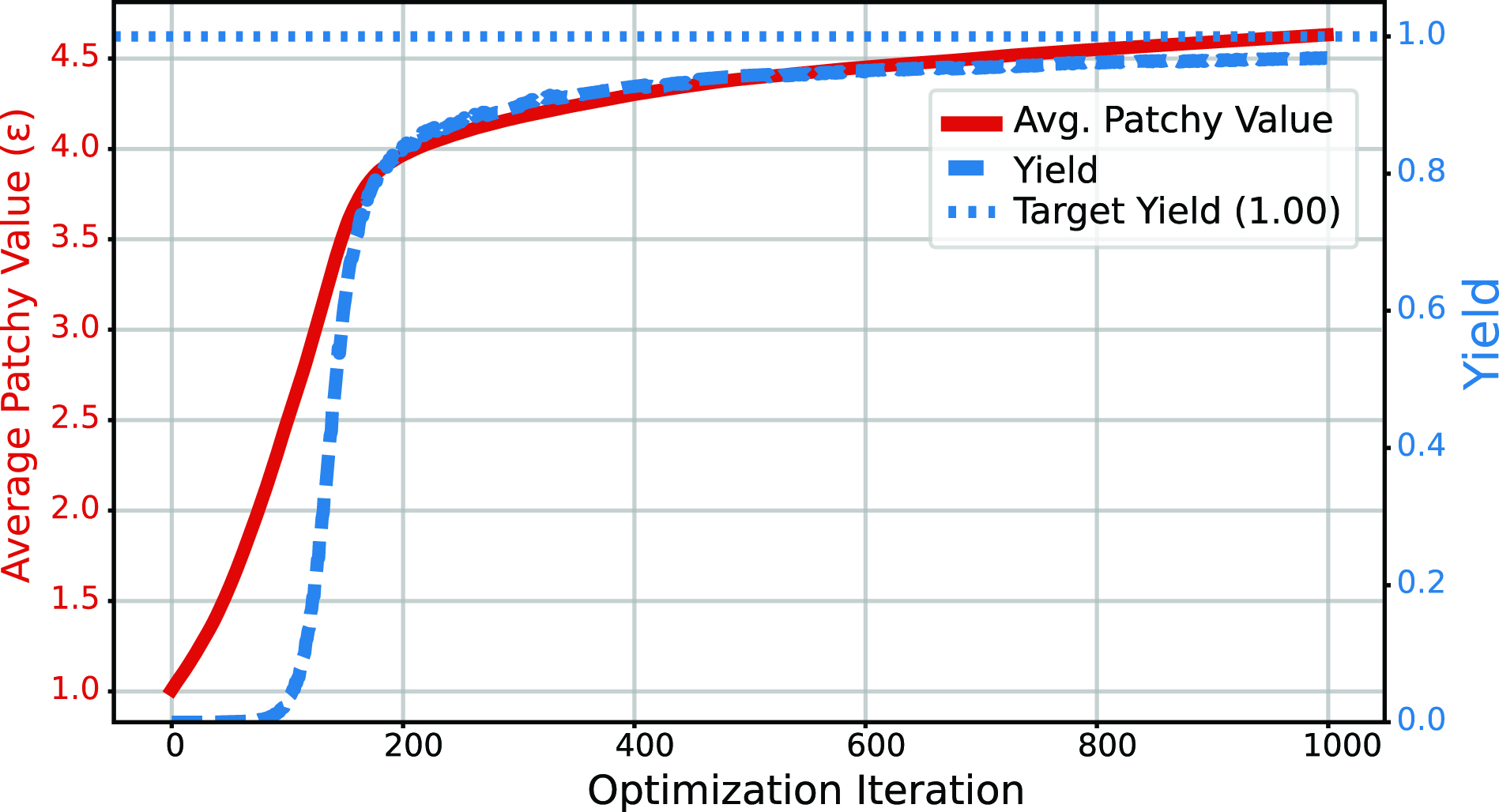}
    \caption{\textbf{Convergence behavior of the optimization with desired yield of 1.0}. The dimer yield fully converges to near 1, value around the 600th iteration. The red curve shows the average attractions strength parameters that are being optimized. after a steep increase curve to $\epsilon_{avg} =4$ the value keeps increasing slowly above 4.5 even after yield convergence has been reached.}
\end{figure}

\begin{figure}[H]
    \centering
    \includegraphics[width=0.7\textwidth]{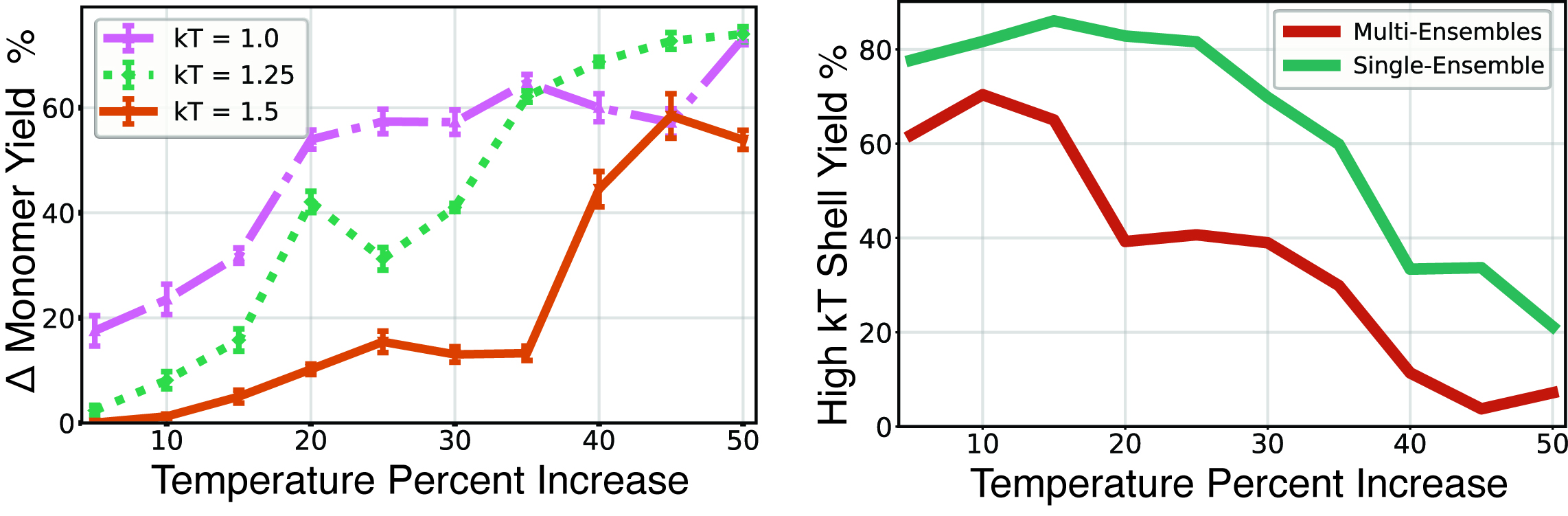}
    \caption{\textbf{Complementary yield trends confirming temperature-dependent assembly behavior.} Because $1-Y_{shell}$ does not directly correspond to the fraction of free monomers -- due to off-target incomplete shells -- these complementary plots confirm the correct trend behavior. Left: Absolute difference in simulated monomer yields between $kT_{low}$ and $kT_{high}$ as a function of the percentage increase in temperature. These trends complement those in Fig. 3C (left), where shell-yield differences were shown instead.
    Right: Simulated shell yields at $kT_{high}$ for the same optimization strategies.}
    \label{fig:opposite_3c}
\end{figure}

\paragraph*{Shell System.}
To capture switch-like behavior for the shell system, the loss is computed over a temperature pair:
\begin{align}
\mathcal{L}_{\text{shell}} = \left|1 - Y_{\text{kT}_{\text{low}}}\right| + \left|Y_{\text{kT}_{\text{high}}}\right| \end{align}
where $Y_{\text{kT}}$ denotes the yield of the shell at temperature $\text{kT}$.
For this objective, we find that extreme parameter initializations (e.g. $\epsilon >> 10$) yield degraded optimization and we therefore initialize parameters using modest interaction strengths for the simulated temperature.
We determined such modest interaction strengths for a given simulation temperature via initial single-ensemble optimizations.
As discussed in the main text, we also find that theoretical yield calculations do not agree with simulated yields with the same precision as in the dimer case (e.g., $>10\%$ relative error).
In the theoretical calculation, the optimization always converges to near-perfect shell assembly at low temperature and near-perfect disassembly at high temperature (i.e., the target behavior), however non-negligible concentrations of intermediate species are also observed upon simulation.
This is likely owing to the large space of candidate off-targets that are not included in the theoretical calculation.
Because the theoretical model employs a truncated state space, the optimizer drives the predicted yield to $\approx 100\%$ within this simplified landscape. While effective for identifying optimal parameters, this theoretical prediction systematically overestimates the yield by neglecting un-enumerated off-targets. We therefore omit these theoretical curves in Figures 3C and 4C, as they would appear as uninformative flat lines relative to the realistic, entropically limited MD results.
Still, the broader trends of switch-like behavior predicted by theory are confirmed with simulation, albeit with less numerical precision.
For the cases in which we compute optimized parameters with the sole intent of maximizing shell yield (used in left panel of Fig. 3C and Fig. \ref{fig:opposite_3c} at a low temperature condition), the loss was defined identically to the dimer case.

\paragraph*{Polymerizing System.}
The polymerizing system introduces a composite loss that balances yield optimization with a penalty term to prevent excessive chain growth. The penalty term is computed independently of any partition function calculation (see SM, Sec.D).
The total loss is defined as:
\begin{align}
\mathcal{L}_{\text{poly}} = \lambda \cdot \left|Y_{\text{target}} - Y_{\text{desired}}\right| + \mathcal{L}_{\text{penalty}},
\end{align}
where $\mathcal{L}_{\text{penalty}}$ is a function of the total predicted concentration of clusters of size $n+1$ (e.g., tetramers; SI D for more detail). 
We use a scaling factor $\lambda$ which we empirically set to 1000.
This loss maintains a stable tradeoff between two intrinsically competing objectives -- suppressing excessive chain growth and promoting high target yield -- while naturally shifting emphasis toward yield once overgrowth is sufficiently minimized and towards $\mathcal{L}_{\text{penalty}}$ when the mass action constraint is significantly violated at the beginning of the optimization process.
To mitigate numerical instability, we clip the monomer concentration and interaction strengths to maintain minimums of $9\times10^{-5}$ and $\varepsilon = 0.25$, respectively.

Optimization trajectories generally converge to a stable plateau within 300 iterations (Fig. S3). For production runs, we report the yield from this sustained stationary segment, which we define as 50 consecutive iterations with minimal variance. Rare late-iteration dips (characterized by abrupt, non-physical yield drops of $\Delta Y\geq0.2$) are identified as numerical artifacts stemming from the projection and clipping steps within the solver; these are consequently excluded from the final reported yields.
\begin{figure}[H]
    \centering
    \includegraphics[width=0.5\textwidth]{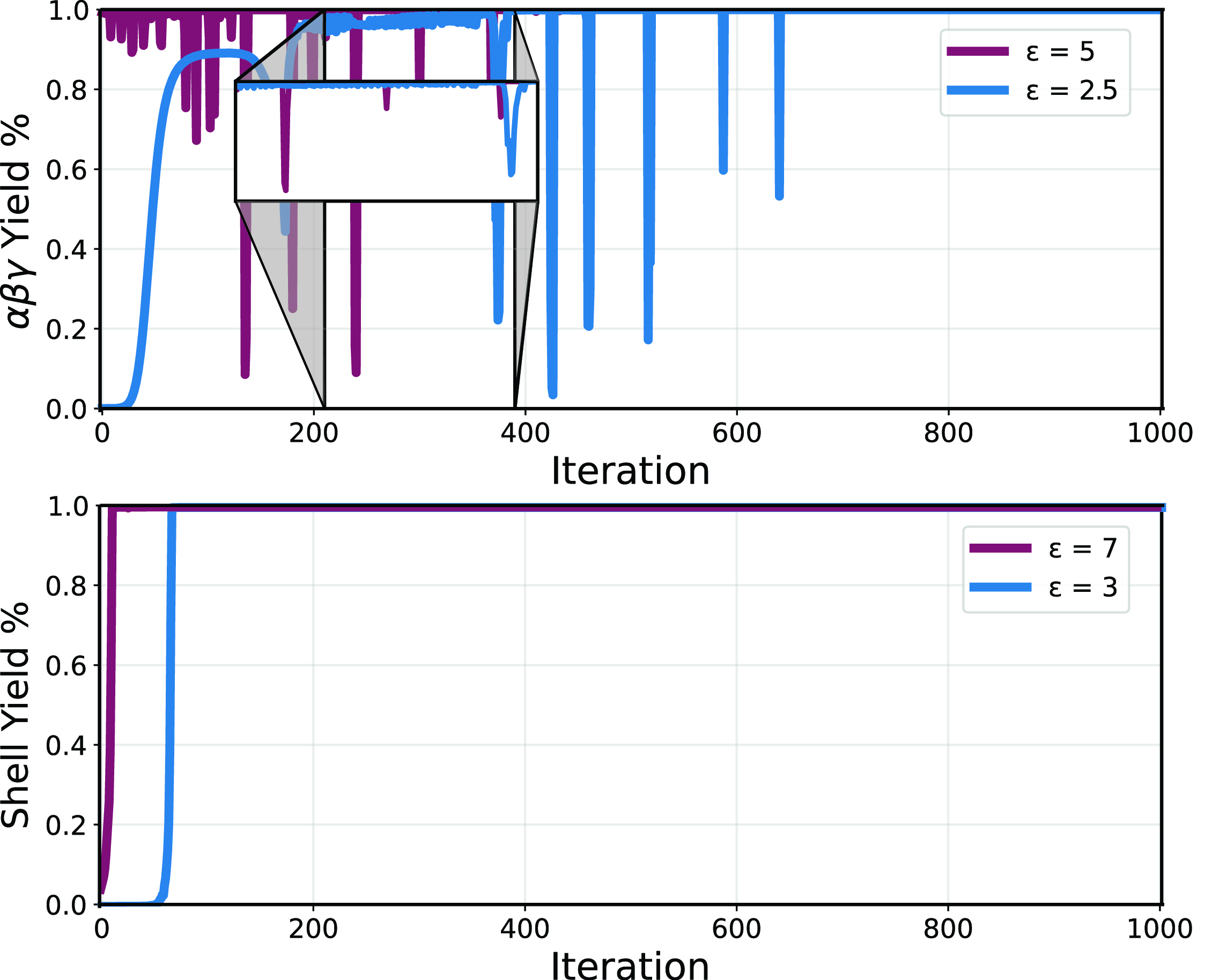}
    \caption{\textbf{Optimization convergence for polymer and shell models.} 
    \textbf{(Top)} Polymer model at $kT = 0.5$ showing the target $\alpha\beta\gamma$ yield over outer-optimization iterations for two initializations of the strong pair-potential attraction: $\varepsilon_{\mathrm{strong}} = 5.0$ (red) and $\varepsilon_{\mathrm{strong}} = 2.5$ (blue). All remaining (``weak'') pair attractions are initialized at $\varepsilon_{\mathrm{weak}} = 2.0$ and optimized concurrently. The inset highlights the convergence window near $\sim 300$ iterations, demonstrating a sustained plateau immediately prior to a rare late-iteration dip. These dips reflect intermittent numerical/optimization excursions (e.g., due to the projection/clipping steps and the sensitivity of the equilibrium solve to parameter updates) rather than a loss of the converged plateau.
    \textbf{(Bottom)} Shell model at fixed $kT = 1.0$ showing shell yield versus iteration for two initial Morse-attraction strengths, $\varepsilon = 7.0$ (red) and $\varepsilon = 3.0$ (blue), with $\varepsilon$ optimized and bounded below by $\varepsilon \ge 0.5$.}
    \label{fig:opt_converge}
\end{figure}

\clearpage
\newpage

\section*{Appendix D: Mass Action Penalty}
In Appendix C, we describe how for the polymerizing system we define a loss function that includes a regularization term to penalize overgrowth. 
 To suppress the optimizer's tendency to converge at parameters which favor unconstrained cluster growth, we turn to classical theories of unbounded self-assembly.

Following the mass action equilibrium expression from Ref.~\cite{hagan2021selflimiting}, the expected concentration $ c_n $ of an $ n $-mer in a single-monomer system is given by:
\begin{equation}
c_n = n \cdot c_1^n \cdot e^{-n\beta \epsilon(n)},
\end{equation}
where $ c_1 $ is the monomer concentration and $ \epsilon(n) $ is the per-subunit free energy of the $ n $-mer relative to $ n $ free monomers.
We extend this model to heterogeneous multi-species systems by defining $ c_1 = c_{\alpha} + c_{\beta} + c_{\gamma} $ and using the structure-level ground state energy $ E_s $ to approximate $ \epsilon(n) $:
\begin{equation}
\epsilon_s(n) = \frac{E_s}{n}.
\end{equation}
Then, to estimate the concentration of a cluster $s$ of size $n$, we apply a generalized multi-species mass action formulation:
\begin{equation}
c^{\mathrm{MA}}_s = n \cdot e^{-\beta \epsilon_s(n)} \prod_{m=1}^{M} c_m^{N_{s,m}},
\label{eq:mass_act_true_supp}
\end{equation}
where $c_m$ is the equilibrium concentration of monomer $m$ obtained from the analytical calculation, $M$ is the number of monomer types, and $ N_{s, m} $ is the count of monomer $m$ in structure $s$. 
Crucially, estimates from Equation \ref{eq:mass_act_true_supp} are computed separately from the yields computed via the procedure introduced by Curatolo et al. \cite{curatolo2023toolbox}, by which we compute full partition functions and map partition functions to yields via a numerical solver.

For the optimizations presented in the main text, we compute the equilibrium concentrations of all clusters up to size $n = 3$ using the calculation of Curatolo et al.
We then compute the overgrowth penalty term for all structures of size $n+1$ via
\begin{align}
\mathcal{L}_{\mathrm{penalty}} &= \eta \cdot \texttt{softplus} \left( \sum_{s \in \mathcal{S}_{n+1}} \log c^{\mathrm{MA}}_s \right), 
\end{align}
where $\mathcal{S}_{n+1}$ denotes the set of all clusters of size $n+1$, and $\eta = 1$ is a tunable scaling factor. 

\clearpage
\newpage

\section*{Appendix E: Simulation Protocols}
As described in the main text, we perform canonical ensemble simulations to validate the parameters obtained via our optimization framework.
Since our optimization targets equilibrium yields, the choice of dynamics does not affect the final thermodynamic ensemble, provided the integrator correctly samples the Boltzmann distribution \cite{frenkel2001}.
To validate the optimized parameters, we employed two distinct simulation environments. For the dimer and polymerizing systems, we utilized HOOMD-blue with the MTTK Nosé-Hoover thermostat \cite{Martyna1994}. 
This method was selected to ensure rigorous sampling of the canonical (NVT) ensemble.
In the case of the shell system, we directly make use of the public simulation code of Krueger et al. which employs the Langevin integration scheme developed by Schoenholz et al. \cite{schoenholz2020jax} implemented JAX-MD.
We note that while experimental self-assembly is typically solvent-mediated and often modeled via Brownian dynamics, we deliberately employed thermostatted Newtonian and Langevin dynamics here to accelerate the exploration of phase space. 
Brownian dynamics, being overdamped, rely on diffusive motion which can be computationally slow to cross energetic barriers. 
In contrast, the inertial terms in Newtonian dynamics allow for ballistic motion, which can significantly reduce the correlation time required to sample independent equilibrium configurations \cite{Leimkuhler2015}.
Since both methods converge to the same canonical distribution, this choice allows for more computationally efficient validation without compromising thermodynamic accuracy.
Looking forward, while optimizing the sampling process itself was not a primary focus of this work, there are clear opportunities for methods development in this area.
Future work could incorporate adaptive stochastic integrators or machine-learning-based samplers, such as Boltzmann Generators \cite{Noe2019}, which have been shown to substantially accelerate convergence to equilibrium for complex many-body systems while preserving the correct thermodynamic ensemble.
In the following sections, we detail the specific simulation parameters and software implementations used for each of the three optimization cases:
\paragraph*{Dimer System.}
We perform canonical ensemble simulations of the dimer system using rigid body dynamics with the MTTK thermostat in the HOOMD-blue (v4.2.1)~\cite{remiHOOMD2024}  molecular dynamics package.
Each simulation contains equal numbers of the two enantiomeric A and B monomers (e.g., $N = 54$ total, with $27$ of each type). 
Monomers are initialized on a simple lattice with one A and one B per unit cell with randomized orientations. 
Periodic boundary conditions are applied. 
We use a timestep of $\Delta t = 10^{-3}$ and a thermostat time constant $\tau = 1.0$ which sets the rate at which the thermostat adjusts the system's kinetic energy to the target temperature.
To match the target concentration, the simulation begins with a box rescaling procedure by which the cubic simulation box is rescaled from the initial lattice using an inverse-volume ramp.
We then apply temperature annealing, beginning from $T = 2.0 + k_BT$ and cooling to the target $k_BT$ in decrements of $0.1$ every $5\times10^{5}$ steps. 
After box rescaling and temperature annealing, we simulate the system for $1.5 \times 10^8$ steps, sampling snapshots every $10^5$ steps.

\paragraph*{Shell System.}
Canonical ensemble simulations of the shell system are performed using JAX-MD (v0.2.8), following the implementation of Krueger et al.~\cite{kruegerShellSim2023}. 
Each simulation is initialized with a periodic cubic box with side length set to achieve a target density of $0.001$. 
Monomer positions are distributed randomly on a cubic lattice with small random displacements, and orientations are sampled uniformly as quaternions. 
The total number of monomers per simulation is $N = 300$.  
Simulations are performed using a Langevin integrator with a timestep of $\Delta t = 10^{-3}$, and a friction coefficient of $\gamma = 1.0$ 
Simulations are run for $2\times10^{7}$ integration steps. 
System states are recorded every $10^{4}$ steps for visualization and analysis.

\paragraph*{Polymerizing Systems.}
The simulation of the polymerizing system largely follows the simulation protocol of the dimer.
However, because we optimize monomer concentrations in continuous space but all simulations are performed in the canonical ensemble, optimized monomer concentrations must be mapped to discrete monomer counts.
To approximate these discrete particle counts from given monomer concentrations, we always simulate a system of $N=300$ monomers, partitioned into the three monomer types according to rounded values from the optimized $\alpha\beta\gamma$ stoichiometry, with a variable side length for the cubic simulation box.
Specifically, at the start of the simulation,
the side length is iteratively adjusted until the resulting overall concentration matches the target value to within a tolerance of $5 \times 10^{-5}$.
All other simulation hyperparameters match those in the dimer case: timestep $\Delta t = 10^{-3}$, thermostat time constant $\tau = 1.0$, and annealing from $T = 2.0 + k_BT$ to the target $k_BT$ in decrements of $0.1$ every $5\times10^{5}$ steps. 
Final production runs are carried out for $6\times10^{8}$ steps, with states sampled every $10^{5}$ steps.

\subsection*{Yield Analysis Methods from Simulation}

For each system and parameter set, three independent simulation replicas are performed with different random seeds, and all reported yields are averaged across these replicas. Yield calculations are based on the final ten frames of each simulation, using particle positions and orientations to identify clusters and determine assembly completeness.

\paragraph*{Dimer System.}
 We identify bonded pairs among core monomers using \texttt{freud.cluster.Cluster} with a cutoff of 2.1. To determine yield, we reconstruct the identity and orientation of each bonded pair and retain only those that form a valid A–B mirror dimer with all patch interactions satisfied. The final dimer yield is defined as the fraction of core monomers that participate in correctly assembled A–B dimers.

\paragraph*{Shell System.}
Connected components are identified using \texttt{freud.cluster.Cluster} with a distance cutoff of 4.2. For each cluster, the number of bonded neighbors per vertex is computed, and a cluster is classified as a complete shell if all six constituent vertices have four bonds, indicating full local connectivity. The total number of shells per frame is then converted into a yield fraction,
$Y_{\text{shell}} = \frac{6\,N_{\text{shell}}}{N_{\text{total}}},
$ representing the fraction of all monomers incorporated into fully assembled shells, averaged over the last ten frames.  To measure disassembly, we use a complementary procedure, where if a monomer has no bond then it classifies as a single monomer cluster.

\paragraph*{Polymerizing System.}
We identify clusters using \texttt{freud.cluster.Cluster} with a distance cutoff of 2.1 and reconstruct the ordered monomer sequence within each cluster via bond-graph traversal. 
All clusters are then categorized by size into monomers, dimers, trimers, and extended chains ($n \ge 4$). 
In the analysis, the yield of off-target trimers is computed as the difference between the total number of trimers and the number of correctly ordered $\alpha\beta\gamma$ trimers. 

\section*{Appendix F: Convergence Analysis}

\begin{figure}[H]
    \centering
    \includegraphics[width=0.9\textwidth]{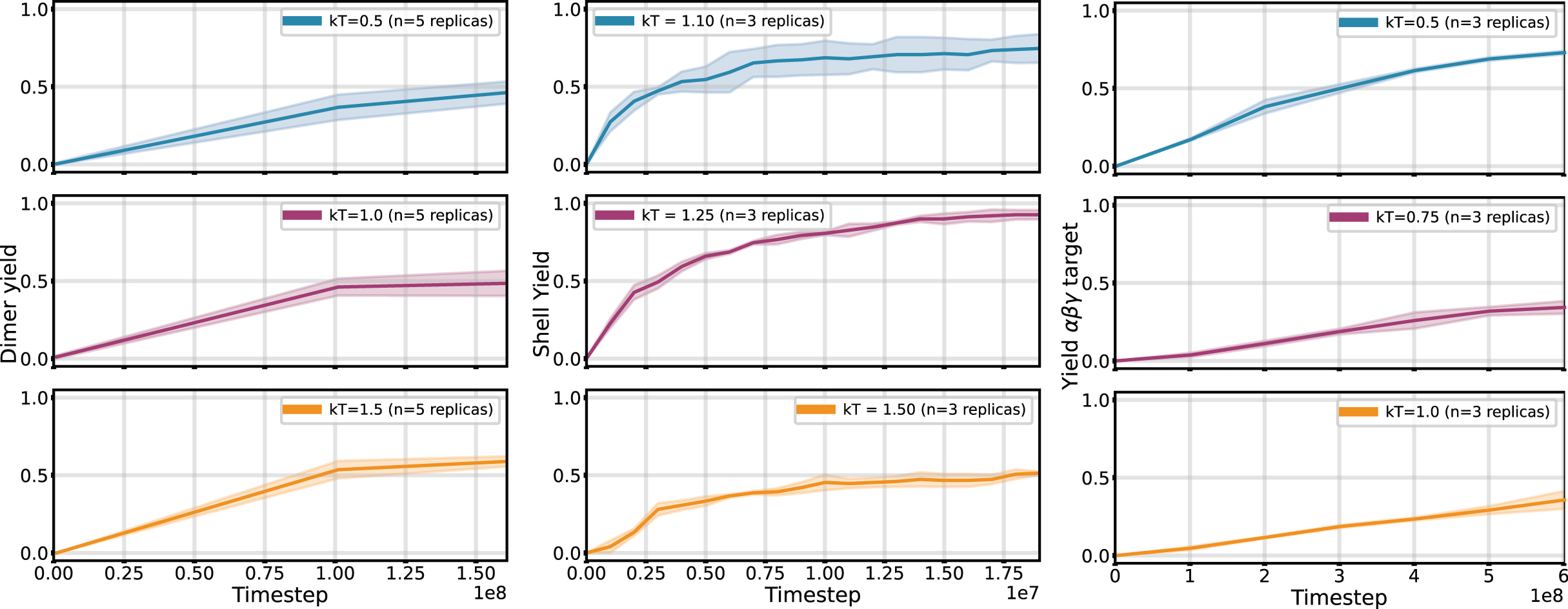}

        \caption{\textbf{Time evolution of assembly yields in molecular dynamics simulations.} 
Representative trajectories of assembly yield as a function of simulation time steps for the Dimer (left column), Shell (middle column), and Polymer (right column) systems. Shaded regions indicate the standard deviation across $n=3$ to $n=5$ independent replicas.
\textbf{(Left)} Dimer yields for parameters optimized to achieve $50\%$ target yield at $kT=0.5, 1.0, 1.5$.
\textbf{(Middle)} Shell yields for parameters optimized at $kT_{low}=1.0$, simulated at increasing temperatures $kT=1.10, 1.25, 1.50$.
\textbf{(Right)} Target $\alpha\beta\gamma$ trimer yields for parameters optimized with mass action regularization at $kT=0.5, 0.75, 1.0$. Note that the polymer simulations were extended to $6 \times 10^8$ steps to assess stability in the dilute limit (see SM, Sec.E for detailed convergence analysis).}
      
    \label{fig:converge}
\end{figure}

To validate that the reported yields represent robust thermodynamic ensembles, we analyzed the time evolution of the assembly yields (Fig.~S4). For the Dimer and Shell systems (Fig.~S4, left and middle panels), the yields exhibit a rapid initial rise followed by a sustained plateau with no significant drift, demonstrating that these systems reached equilibrium within their respective simulation durations ($1.5 \times 10^8$ steps for the dimer and $2 \times 10^7$ steps for the shell).
For the Polymer system (Fig.~S4, right panel), the assembly kinetics are governed by diffusion-limited aggregation due to the low monomer concentration ($c_{tot}=10^{-4}$) and the entropic barrier associated with trimer nucleation. In this dilute regime, the mean free path between monomers is large, and the timescale required for exhaustive sampling of the equilibrium distribution significantly exceeds standard molecular dynamics timescales \cite{frenkel2001, whitelam2015}.
To capture these slower kinetics, we performed production runs of $6 \times 10^8$ steps for the polymer system. As shown in Fig.~S4 (right panel), the trajectories exhibit a sustained positive drift throughout the simulation, indicating that the system is slowly approaching equilibrium from below. This continuous increase in yield confirms that the optimized parameters have successfully created a thermodynamic basin for the target structure, even though kinetic constraints limit the rate of assembly. Consequently, the yields reported in the main text should be interpreted as \textit{conservative lower bounds} of the true equilibrium values, limited only by the finite timescales of diffusion in dilute media \cite{hagan2021selflimiting}.
\newpage
\clearpage

\bibliographystyle{apsrev4-2}
\bibliography{references}

\end{document}